\documentclass[12pt]{spieman}  
\usepackage{amsmath,amsfonts,amssymb}
\usepackage{graphicx}
\usepackage{setspace}
\usepackage{tocloft}
\usepackage{lineno}
\usepackage{booktabs}
\usepackage{multirow} 
\usepackage{caption} 
\usepackage{subcaption}
\usepackage[numbers,sort&compress]{natbib}

\title{Unpaired Deep Learning Synthesis of Photon-Counting CT Material Basis Maps from Non-contrast Energy-Integrating Abdominal CT Images }

\author[a,b,*]{Ruihan Huang}
\author[c,d]{Staffan Holmin}
\author[a,b,e]{Mats Persson}
\affil[a]{KTH Royal Institute of Technology, Department of Physics, SE-10691, Stockholm, Sweden}
\affil[b]{Karolinska University Hospital,  MedTechLabs, BioClinicum, Solna, Sweden}
\affil[c]{Karolinska Institutet,  Department of Clinical Neuroscience, Stockholm, Sweden}
\affil[d]{Karolinska University Hospital,  Department of Neuroradiology, Stockholm, Sweden}
\affil[e]{Digital Futures, KTH Royal Institute of Technology,   Stockholm, Sweden}

\cftpagenumbersoff{figure}
\cftpagenumbersoff{table} 
\begin{document} 
\maketitle

\begin{abstract}

\textbf{Purpose:} Photon-counting Computed Tomography (PCCT) is the most advanced Computed Tomography (CT) technology, offering significant improvements in image quality and diagnostic capabilities. However, since PCCT has only recently been adopted in the clinc, there are no publicly available PCCT image datasets for study. We therefore aim to synthesize PCCT spectral material-basis images from publicly available EID CT images.

\textbf{Approach:} We propose a two-step deep learning model designed to synthesize photon-counting spectral material basis images from public Energy-Integrating Detector (EID) CT images. In the first step, we use a Denoising Diffusion Implicit Model (DDIM) to generate EID CT images from PCCT images. In the second step we use a U-Net with a Domain-Adversarial Neural Network to predict water and iodine maps from generated EID CT images. We also reconstruct basis images and virtual monoenergetic images (VMIs) from the predicted material-basis maps for evaluation.

\textbf{Results:} We evaluated the generated water and iodine maps as well as the 40 and 70 keV PCCT images in terms of Hounsfield Unit accuracy, modulation transfer function and noise power spectrum as well as qualitative image appearance. The reconstructed 40 and 70 keV PCCT images exhibit higher spatial resolution while preserving the anatomical structures and textures of the original EID CT images, thereby demonstrating the feasibility of the proposed approach.

\textbf{Conclusions:} The proposed framework provides a feasible approach for synthesizing PCCT spectral material-basis images from conventional EID CT without requiring paired images. This method has the potential to provide large sets of synthetic training and evaluation data for PCCT algorithm development in data-limited environments.
\end{abstract}

\begin{enumerate}
    \item 
\end{enumerate}

{\noindent \footnotesize\textbf{*}Ruihan Huang,  \linkable{ruihanh@ug.kth.se} }

\begin{spacing}{2}   

\section{Introduction}
\label{sect:intro}  
Photon-counting computed tomography (PCCT) is a state-of-the-art technology in CT imaging. Compared with conventional energy-integrating detector (EID) CT, photon-counting spectral (energy-resolving) CT can reduce detector electronic noise, improve dose efficiency, and enhance spatial and contrast resolution, resulting in improved image quality while enabling radiation-dose reduction \cite{Almqvist2024SecondGenSiPCCT, Danielsson2021PhotonCountingDetectorsCT}. In addition, PCCT inherently provides spectral information that supports material decomposition (e.g., water/iodine), whereas EID CT does not directly output these basis images \cite{Danielsson2021PhotonCountingDetectorsCT}. Recent research on PCCT is rapidly expanding, with substantial  work on image denoising, artifact reduction, and motion-correction tasks that typically require large volumes of clinical PCCT data for training, validation, and benchmarking. In contrast to EICT, for which several large datasets are available, PCCT has only recently entered clinical practice and publicly available clinical PCCT datasets remain scarce, which restricts research. Although CT simulation tools such as CatSim \cite{DeMan2007CatSim} can generate PCCT-like images based on numerical phantoms, phantom-based simulations often fail to capture the full complexity and variability of real clinical anatomy, which can limit their usefulness for some research applications \cite{huang2026pcct_motion_correction}. Therefore, providing access to realistic PCCT images is a critical challenge.

Medical image synthesis offers a practical solution to this challenge. Medical image synthesis refers to using algorithms or learned models to generate realistic images or to translate images between modalities, thereby supporting downstream clinical and computational workflows. Conventional synthesis approaches include bulk density assignment and atlas-based methods (often used to produce pseudo-CT images via registration to a reference atlas) \cite{Johnstone2018SystematicReviewSyntheticCT}. Although these classical methods are typically interpretable and can perform well under controlled assumptions, they often struggle to capture complex, nonlinear cross-modality relationships. Recently, the demonstrated success of deep learning for core medical imaging tasks, including segmentation, lesion and structure detection, and CT image reconstruction, has motivated increasing research into deep-learning based medical image synthesis. The models are commonly used either to approximate the underlying data distribution or to learn conditional mappings that preserve anatomical structure while producing a target-modality appearance that is physically and statistically consistent\cite{Nie2017ContextAwareGAN, Wolterink2017UnpairedMRtoCT}, such as translating CT images into an MRI-like appearance\cite{Lyu2022CTMRIDiffusion}.

Most studies of deep-learning-based medical image synthesis have focused on generative adversarial networks (GANs). In a typical GAN framework, a generator is trained to produce synthetic images, while a discriminator learns to distinguish real images from generated ones, thereby encouraging realistic outputs \cite{Yi2019GANMedicalImagingReview}. For example, Nie et al. proposed a context-aware GAN for MRI→CT synthesis, showing that adversarial training can yield CT images with more realistic appearance than purely regression-based methods\cite{Nie2017ContextAwareGAN}. Since paired multi-modal medical images are often unavailable or imperfectly aligned, Wolterink et al. demonstrated GAN-based MR→CT synthesis using unpaired data, enabling modality translation without requiring paired training scans \cite{Wolterink2017UnpairedMRtoCT}. 

More recently, diffusion models have attracted increasing attention for medical image synthesis due to their strong ability to model complex image appearance and noise characteristics. Pan et al. proposed an improved denoising diffusion probabilistic model for MRI-conditioned synthetic CT generation, reporting high-quality results for this translation task \cite{Pan2024TransformerDiffusionSyntheticCT}. Lyu et al. adapted diffusion and score-matching models for CT to MRI conversion and reported improved synthetic image quality compared with CNN/GAN baselines in their experiments\cite{Lyu2022CTMRIDiffusion}. Although several studies have investigated deep-learning synthesis for CT and spectral CT, most of this work has focused on dual-energy CT (DECT) rather than photon-counting CT\cite{Xu2024WaveletLossCycleGANsVMI, Cong2020VirtualMonoenergeticCTPatterns, Kawahara2021MonoenergeticCTGAN, Lyu2021EstimatingDualEnergyCT, Li2023QualityCheckedPhysicsConstrained, Li2026ImageQualityAssessmentDLVME}. These methods commonly rely on paired or corresponding training data, where single-energy CT images are associated with reference DECT-derived images, material-basis maps, or VMIs.To the best of our knowledge, there is a lack of research specifically addressing the generation of PCCT images from EID CT data. Recently, Liu et al. proposed SUMI, a degradation-modeling framework that distills photon-counting CT image characteristics into routine chest EID CT images by simulating clinically validated degradations from PCCT and learning to reverse them \cite{Liu2026SUMI}. Their work demonstrates the potential of using limited high-quality PCCT data to improve conventional EID CT images and supports the broader feasibility of PCCT-like image synthesis. However, this approach primarily focuses on enhancing routine CT image quality, whereas the synthesis of PCCT spectral material-basis information, such as water and iodine maps, from conventional EID CT remains less explored. In contrast, our study aims to generate PCCT-derived material-basis maps and VMIs from EID CT images without requiring paired EID CT-PCCT training data.

The purpose of this work is to provide a method for the synthesis of PCCT material-basis images from single-energy EID-CT images based on only a small set of PCCT training data. Since paired PCCT and EID CT images are not available, we decompose the overall problem into two steps. First, we train a diffusion model on real EID CT images to capture the appearance distribution of EID CT, and then use it to generate EID-like CT images based on PCCT images, constructing a pseudo-paired dataset for the second step. Next, we train a supervised model to predict PCCT material-basis images from the synthesized EID CT images. Finally, we demonstrate the feasibility of the proposed approach and evaluate its performance using both qualitative visual assessments and quantitative metrics.

\section{Materials and Methods}

\subsection{Data}\label{subsec2}

The EID CT images used in this study were abdominal-region slices selected from  the Lung CT Segmentation Challenge 2017 (LCTSC) dataset \cite{Yang2017LCTSC}. The abdomen PCCT images and material basis images (iodine and water maps) are obtained retrospectively from a silicon-based photon counting CT prototype (GE Healthcare) at Karolinska Institute, acquired with approval from the Swedish Ethics Review Authority (permits 2023-03709-01, 2024-02038-02, and 2024-07310-02)\cite{Almqvist2024InitialClinicalImagesSiPCCT}. The dataset comprised 2500 EID CT slices from 30 patients, and 2500 PCCT slices comprising material-basis map pairs from 8 patients. From these data, 2200 images were randomly selected for training, 100 images for validation, and 200 images for testing.  To improve training stability, we performed image registration in MATLAB \cite{matlab_r2023b} to align EID CT and PCCT images to a common template. Because all PCCT slices are acquired by the same PCCT system and detector, they are already intrinsically aligned in a common scanner coordinate system and therefore do not require additional inter-slice registration. In contrast, the EID CT images come from a different acquisition domain, so we registered only the EID CT images to a fixed PCCT template slice to reduce cross-domain misalignment and ensure consistent anatomical correspondence for training. Specifically, we selected one PCCT image from the dataset as the fixed reference (template) and treated each EID CT image as the moving image.We then performed intensity-based multimodal registration using the \texttt{imregtform} function  \cite{MathWorks_imregtform_doc} with a multimodal configuration and a mutual-information similarity metric. The transformation model was set to \texttt{similarity}, which estimates a global rigid transformation with isotropic scaling, including translation, rotation, and uniform scaling. No deformable or non-rigid registration was applied. We further applied horizontal and vertical flipping for data augmentation. For preprocessing, all EID CT and PCCT images were clipped to a fixed Hounsfield unit (HU) window of [$-1000$, 2000] and then normalized to [$-1$, 1] using global min–max scaling. The PCCT material-basis images (iodine and water maps) were normalized to [$-1$, 1] using global min–max scaling, where the minimum and maximum values were computed over the entire set of material maps.

\subsection{Networks}\label{subsec2}

Since paired EID CT and PCCT images were not available, and because directly mapping a single EID CT image to two material-basis images (water and iodine) is a challenging and highly constrained task, we formulated the material-basis prediction as a supervised learning problem and decomposed the overall pipeline into two stages. First, we applied an unsupervised (unpaired) domain translation approach to map PCCT images into the EID CT domain, thereby generating EID-like images from PCCT scans. This step enabled us to construct a pseudo-paired dataset consisting of synthesized EID CT images and the corresponding PCCT material-basis images. Second, we trained a supervised model on this pseudo-paired dataset to learn the mapping from EID CT images to PCCT material-basis images.
\subsubsection{Step 1: pseudo-paired dataset generation}\label{subsubsec2}
For the first stage, we adopt Denoising Diffusion Implicit Models (DDIM) \cite{song2022denoisingdiffusionimplicitmodels} to synthesize EID CT images. Specifically, we train a DDIM model on real EID CT images so that the network learns the EID CT data distribution (including characteristic appearance and noise texture). DDIM can be viewed as a fast-sampling formulation derived from Denoising Diffusion Probabilistic Models (DDPM) \cite{ho2020ddpm}. DDPM generation reverses a Markovian Gaussian diffusion process and often requires hundreds to thousands of sampling steps, whereas DDIM defines a non-Markovian process with the same per-timestep marginals, enabling the same training objective but allowing substantially fewer sampling steps during inference. The forward diffusion process is:\begin{equation}
x_t=\sqrt{\bar{\alpha}_t}\,x_0+\sqrt{1-\bar{\alpha}_t}\,\epsilon,
\qquad \epsilon\sim\mathcal{N}(0,I).
\end{equation}
The loss function is:
\begin{equation}
\mathcal{L}_{\text{simple}}(\theta)
= \mathbb{E}_{\mathbf{x}_0,\, t,\, \boldsymbol{\epsilon}}
\left[\left\| \boldsymbol{\epsilon}
- \boldsymbol{\epsilon}_\theta(\mathbf{x}_t, t)\right\|_2^2\right].
\end{equation}
where $t\sim\mathrm{Uniform}\{1,\dots,T\}$ and $\epsilon_\theta(x_t,t)$ denotes the network prediction of the added noise. In this work, we used a U-Net \cite{ronneberger2015unetconvolutionalnetworksbiomedical} as the noise-prediction network.
During sampling, DDIM uses the same trained network but allows a coarser timestep schedule.

After training, we inject a PCCT image into the reverse diffusion process to generate an EID-like output. Concretely, at a chosen timestep $t_h$, we replace the current latent with a noised PCCT image obtained by applying h steps of the forward diffusion process, and then continue DDIM sampling. To optimize the performance of Step 1, we introduce a noise-strength control for the hijack operation. Let $x_0$ denote the PCCT image and let $x_H=\sqrt{\bar{\alpha}_H}\,x_0+\sqrt{1-\bar{\alpha}_H}\,\epsilon$ be the standard forward-diffused sample at step $H$, with $\epsilon\sim\mathcal{N}(0,I)$.
We then construct a partially noised image,
\begin{equation}
\tilde{x}_H = x_0 + \gamma\,(x_H-x_0) = (1-\gamma)x_0 + \gamma x_H .
\end{equation}
where $\gamma\in[0,1]$ controls the corruption strength (higher $\gamma$ yields stronger noising).

Starting from $\tilde{x}_H$, we perform DDIM reverse-time sampling for $t=H,\dots,1$.
At each step, we first estimate
\begin{equation}
\hat{x}_0(\tilde{x}_t,t)=
\frac{1}{\sqrt{\bar{\alpha}_t}}
\left(\tilde{x}_t-\sqrt{1-\bar{\alpha}_t}\,\epsilon_\theta(\tilde{x}_t,t)\right),
\end{equation}
and then update (deterministic DDIM, i.e., $\eta=0$)
\begin{equation}
\tilde{x}_{t-1}=
\sqrt{\bar{\alpha}_{t-1}}\,\hat{x}_0(\tilde{x}_t,t)
+\sqrt{1-\bar{\alpha}_{t-1}}\,\epsilon_\theta(\tilde{x}_t,t),
\end{equation}
where $\epsilon_\theta(\cdot,t)$ is the trained noise-prediction network.
The final output $\tilde{x}_0$ is taken as the synthesized EID-like CT image.
This “hijack” operation effectively uses the learned EID prior while preserving the PCCT content, producing synthesized EID CT images that can be paired with the corresponding PCCT material-basis maps for the second-stage supervised training.

During the training for Step 1, the diffusion network was trained using a DDIM sampling scheme with 200 timesteps. The network was trained for 300,000 iterations using image patches of size $512 \times 512$ pixels and a batch size of 4.

\subsubsection{Step 2: mapping from EID CT images to PCCT material-basis maps}\label{subsubsec2}

For Step 2, we used U-Net \cite{ronneberger2015unetconvolutionalnetworksbiomedical} as the material-prediction network to estimate the water and iodine basis maps from the synthetic EID CT images generated in Step 1. U-Net is a widely used encoder–decoder model with skip connections that supports accurate, spatially resolved prediction in biomedical imaging tasks. 

Since the synthetic EID CT images are not perfectly identical to real EID CT images, a domain shift exists between the Step-1 outputs (synthetic domain) and the real EID CT images (target domain). To improve robustness under this shift, we augment U-Net with a Domain-Adversarial Neural Network (DANN) module \cite{ganin2016domain}, which encourages the learned features of the generator (UNet in our work) to be informative for the main task but uninformative about the domain. DANN achieves this by adding an auxiliary domain classifier connected through a Gradient Reversal Layer (GRL): the domain classifier is trained to distinguish synthetic vs. real features, while the GRL reverses (and scales) the gradient so that the feature extractor learns to confuse the domain classifier.

Let $G_\theta$ denote the U-Net predictor. Given an input CT image $x$, the network outputs
\begin{equation}
(\hat{W},\hat{I}) = G_\theta(x),
\end{equation}
where $\hat{W}$ and $\hat{I}$ are the predicted water and iodine maps, and $(W,I)$ are the corresponding ground-truth maps (available for paired source data).
Let $\phi_j(\cdot)$ denote the activation at layer $j$ of a VGG16 network\cite{Simonyan2015VGG} pretrained on ImageNet\cite{Deng2009ImageNet} and $\mathcal{J}$ the set of selected layers.
We define the combined water and iodine losses as
\begin{align}
\mathcal{L}_{W} &= \lambda_{\ell_1}\|\hat{W}-W\|_1 \;+\; \lambda_{\mathrm{VGG}}\sum_{j\in\mathcal{J}}\left\|\phi_j(\hat{W})-\phi_j(W)\right\|_1, \\
\mathcal{L}_{I} &= \lambda_{\ell_1} \|\hat{I}-I\|_1 \;+\; \lambda_{\mathrm{VGG}}\sum_{j\in\mathcal{J}}\left\|\phi_j(\hat{I})-\phi_j(I)\right\|_1 ,
\end{align}
where the VGG term uses an $L_1$ distance in feature space. The $\lambda_{\ell_1}$ and $\lambda_{vgg}$ denote the weighting factors of the $\ell_1$ loss and the VGG perceptual loss, respectively. 
To balance the two terms, we set $(\lambda_{\ell_1},\lambda_{\mathrm{VGG}})=(1,10)$ for the water-map loss $\mathcal{L}_W$ and $(\lambda_{\ell_1},\lambda_{\mathrm{VGG}})=(1,20)$ for the iodine-map loss $\mathcal{L}_I$ in our experiments.

Let $f_\theta(\cdot)$ be the feature representation and $D_\psi(\cdot)$ a domain classifier. In our implementation, 
$D_\psi(\cdot)$ is a multilayer perceptron (MLP)\cite{goodfellow2016deep}, which is a common design choice in DANN-style domain-adversarial learning\cite{ganin2016domain}.
Using a gradient reversal layer (GRL), the domain-adversarial loss is
\begin{equation}
\mathcal{L}_{\mathrm{DANN}} =
\mathbb{E}_{x_s}\!\left[\mathrm{CE}\!\left(D_\psi(\mathrm{GRL}(f_\theta(x_s))),\,0\right)\right]
+
\mathbb{E}_{x_t}\!\left[\mathrm{CE}\!\left(D_\psi(\mathrm{GRL}(f_\theta(x_t))),\,1\right)\right],
\end{equation}
where $x_s$ and $x_t$ denote source (synthetic EID CT) and target (real EID CT) inputs, respectively.

\begin{equation}
\mathcal{L}_{\mathrm{step2}} =
\lambda_{\mathrm{DANN}}\mathcal{L}_{\mathrm{DANN}}
+ \lambda_{W}\mathcal{L}_{W}
+ \lambda_{I}\mathcal{L}_{I}.
\end{equation}
where $\lambda$ denotes the weighting factor for each loss term.

We used a U-Net architecture with a one-channel input and two-channel output and 64 base feature channels. Our U-Net has 4 encoder and 4 decoder levels (5 resolution scales in total, including the bottleneck), with 19 convolutional layers in total (8 in the encoder, 2 in the bottleneck, 8 in the decoder, and a final 1×1 output convolution). For the DANN module, multi-level encoder features were extracted from the first three downsampling blocks of the U-Net, denoted as \texttt{down1}, \texttt{down2}, and \texttt{down3}. Each feature level was connected to an independent patch-based domain discriminator through a gradient reversal layer. The GRL coefficient was progressively increased during training according to the DANN scheduling function, with a maximum scaling factor of 2.0. The network was implemented in PyTorch~\cite{paszke2019pytorch} and trained using the Adam optimizer~\cite{kingma2015adam} with a learning rate of $1\times10^{-4}$, batch size 2, and 100 training epochs. All experiments were run on a workstation equipped with an NVIDIA RTX A6000 GPU (48 GB memory).

\subsection{Evaluation methods}\label{subsec2}
\subsubsection{Noise power spectrum}\label{subsubsec2}

Because Step 1 uses unpaired PCCT and EID CT data, pixel-by-pixel error metrics are not meaningful. Instead, we evaluated noise characteristics using the noise power spectrum (NPS) \cite{Riederer1978NPS} computed from anatomically matched ROIs. NPS quantifies the spatial-frequency content of image noise and therefore captures not only the overall noise magnitude but also noise texture and correlation, which are key differences between PCCT and EID CT images. We selected 50 slices from the EID CT, PCCT, and synthesized EID CT datasets that have the same organ (liver) at the same position, and computed the NPS within selected ROIs. For each image, a $60 \times 60$ pixel ROI was selected at the same image location. 
For a two-dimensional ROI $I(x,y)$ of size $N_x \times N_y$, the mean-subtracted noise image was defined as
\begin{equation}
\Delta I(x,y) = I(x,y) - \bar{I},
\end{equation}
where $\bar{I}$ denotes the mean intensity within the ROI. The two-dimensional NPS was then computed as
\begin{equation}
\mathrm{NPS}(f_x,f_y) =
\frac{1}{N_x N_y}
\left|
\mathcal{F}\left\{\Delta I(x,y)\right\}
\right|^2,
\end{equation}
where $\mathcal{F}\{\cdot\}$ denotes the two-dimensional Fourier transform, and $f_x$ and $f_y$ are the spatial frequencies along the horizontal and vertical directions, respectively. The frequency axis was normalized in units of cycles/pixel and ranged from 0 to the Nyquist frequency of 0.5 cycles/pixel. The final one-dimensional NPS curve was obtained by radially averaging the two-dimensional NPS over annular frequency bins:
\begin{equation}
\mathrm{NPS}(f) =
\left\langle
\mathrm{NPS}(f_x,f_y)
\right\rangle_{\sqrt{f_x^2+f_y^2}=f}.
\end{equation}

\subsubsection{Material basis algorithm}\label{subsubsec2}

For Step 2, we evaluated not only the predicted material-basis images (water and iodine maps) but also the derived virtual monoenergetic images (VMIs). Since VMIs are often of greater clinical interest than basis images, we computed VMIs from the predicted material maps and performed quantitative evaluation at 40 and 70 keV VMIs.

To generate VMIs, we use the fact that the linear attenuation coefficient at a monoenergetic level \textit{E} can be approximated as a linear combination of the selected basis materials (water and iodine).
\begin{equation}
    \mu(\mathbf{r},E) \approx w(\mathbf{r})\,\mu_{\text{water}}(E) + i(\mathbf{r})\,\mu_{\text{iodine}}(E),
\end{equation}
where $w(r)$ and $i(r)$ are the water and iodine basis coefficients, and $\mu_{\text{water}}(E)$, $\mu_{\text{iodine}}(E)$ are the basis functions of water and iodine at energy $E$.

\subsubsection{Quantitative evaluation method for validation dataset}\label{subsubsec2}

We evaluated the Step 2 predictions on the validation set using peak signal-to-noise ratio (PSNR)\cite{huynhthu2008psnr}, structural similarity index (SSIM)\cite{wang2004ssim}, and Learned Perceptual Image Patch Similarity (LPIPS)\cite{zhang2018lpips}, because this set provides paired data (synthetic EID CT inputs with corresponding ground-truth spectral outputs). 

PSNR measures pixel-wise fidelity via the mean squared error (MSE).
\begin{equation}
\mathrm{MSE}(x,\hat{x})=\frac{1}{N}\sum_{i=1}^{N}(x_i-\hat{x}_i)^2,
\end{equation}
\begin{equation}
\mathrm{PSNR}(x,\hat{x})=10\log_{10}\left(\frac{L^2}{\mathrm{MSE}(x,\hat{x})}\right),
\end{equation}
where $x$ and $\hat{x}$ denote the reference and predicted images, respectively, $N$ is the total number of pixels/voxels, and $L$ is the image dynamic range (e.g., $L=2^{B}-1$ for $B$-bit images). Higher PSNR indicates closer agreement.

SSIM evaluates perceived similarity by comparing local luminance, contrast, and structure; SSIM ranges (typically) from [$-1$, 1] with 
1 being perfect similarity.
\begin{equation}
\mathrm{SSIM}(x,y)=
\frac{(2\mu_x\mu_y+c_1)(2\sigma_{xy}+c_2)}
     {(\mu_x^2+\mu_y^2+c_1)(\sigma_x^2+\sigma_y^2+c_2)},
\end{equation}
where $\mu_x$ and $\mu_y$ are the means, $\sigma_x^2$ and $\sigma_y^2$ are the variances, and $\sigma_{xy}$ is the covariance. The constants $c_1=(k_1L)^2$ and $c_2=(k_2L)^2$ are used to stabilize the division, where $L$ is the dynamic range of the pixel values. We used $k_1=0.01$ and $k_2=0.03$ as recommended in the original publication\cite{wang2004ssim}.

LPIPS measures perceptual similarity using deep network features. It computes a weighted distance between normalized feature extracted from a pretrained network, and we adopt AlexNet\cite{Krizhevsky2012AlexNet} as the feature extractor in our experiments. Lower LPIPS indicates higher perceptual similarity.

Let $F$ be a pretrained network, and let $\hat{y}^{\,l},\,\hat{y}^{\,l}_0 \in \mathbb{R}^{H_l \times W_l \times C_l}$
denote the channel-wise normalized feature maps at layer $l$ for images $x$ and $x_0$, respectively.
LPIPS is defined as
\begin{equation}
d(x,x_0)=\sum_{l}\frac{1}{H_lW_l}\sum_{h,w}
\left\|\, w_l \odot \left(\hat{y}^{\,l}_{hw}-\hat{y}^{\,l}_{0\,hw}\right)\right\|_2^2,
\end{equation}
where $w_l$ are learned channel-wise weights and $\odot$ denotes element-wise multiplication.

\subsubsection{Quantitative evaluation method for tests dataset}\label{subsubsec2}

For the test set (real EID CT images), we used modality independent neighbourhood descriptor (MIND)\cite{Heinrich2012MIND} to assess how well the network preserves anatomical structure when paired ground truth is unavailable. MIND is a modality-independent descriptor based on local self-similarity patterns that tend to be stable across imaging domains; therefore, a smaller MIND distance between the generated images and the input EID CT suggests better structural preservation under domain shift.
MIND at voxel/location $x$ with offset $r\in R$ is defined as
\begin{equation}
\mathrm{MIND}(I;x;r)=\frac{1}{n}\exp\!\left(
-\frac{D_p(I;x,x+r)}{V(I;x)}
\right),\qquad r\in R,
\end{equation}
where $n$ is a normalizing constant such that the maximum descriptor value equals $1$.

The patch distance $D_p$ was computed as the Gaussian-weighted sum of squared differences (SSD) between local patches $P$ centered at two locations:
\begin{equation}
D_p(I;x_1,x_2)
=
\sum_{p\in P}
G_{\sigma}(p)
\left(I(x_1+p)-I(x_2+p)\right)^2,
\end{equation}
where $G_{\sigma}$ denotes a Gaussian weighting kernel. In our implementation, $\sigma=0.5$ was used, corresponding to an effective patch size of $3 \times 3$ pixels.

The local variance term was computed as the mean of patch distances within the four-neighborhood $N$ in the 2D image plane:
\begin{equation}
V(I;x)=\frac{1}{|N|}\sum_{n\in N}D_p(I;x,x+n),
\end{equation}
where $N=\{(1,0),(-1,0),(0,1),(0,-1)\}$ and $|N|=4$.

To compare two images $I$ and $\hat{I}$ structurally, we computed the mean squared difference between their MIND descriptors over the image domain $\Omega$:
\begin{equation}
\mathcal{D}_{\mathrm{MIND}}(I,\hat{I})=
\frac{1}{|\Omega||R|}\sum_{x\in\Omega}\sum_{r\in R}
\left(\mathrm{MIND}(I;x;r)-\mathrm{MIND}(\hat{I};x;r)\right)^2.
\end{equation}

To evaluate the spatial resolution of the generated images, we also calculated and compared the modulation transfer function (MTF) between the real EID CT images and the corresponding generated PCCT-derived images. The MTF characterizes the ability of an imaging system to preserve object contrast as a function of spatial frequency and is commonly used to quantify spatial resolution. In this study, the MTF was estimated using an edge-based method. For each manually selected ROI, the edge spread function (ESF) was obtained by averaging the line profiles perpendicular to the dominant edge. The ESF was interpolated using sub-pixel sampling, differentiated to obtain the line spread function (LSF). The MTF was estimated as the normalized magnitude of the Fourier transform of the LSF \cite{Samei1998PresampledMTF}:

\begin{equation}
\mathrm{MTF}(f) =
\frac{\left| \mathcal{F}\{\mathrm{LSF}(x)\}(f) \right|}
{\left| \mathcal{F}\{\mathrm{LSF}(x)\}(0) \right|},
\end{equation}
where $f$ denotes the spatial frequency, $\mathcal{F}\{\cdot\}$ represents the Fourier transform, and $\mathrm{LSF}(x)$ is the line spread function derived from the edge spread function. A higher MTF value at a given spatial frequency indicates better preservation of fine structural details. Comparison of the MTF curves provides a quantitative assessment of whether the generated images improve or preserve spatial resolution relative to the original EID CT images.

\section{Results}\label{sec3}

\begin{figure}
\begin{center}
\begin{tabular}{c}
\includegraphics[height=10.0cm]{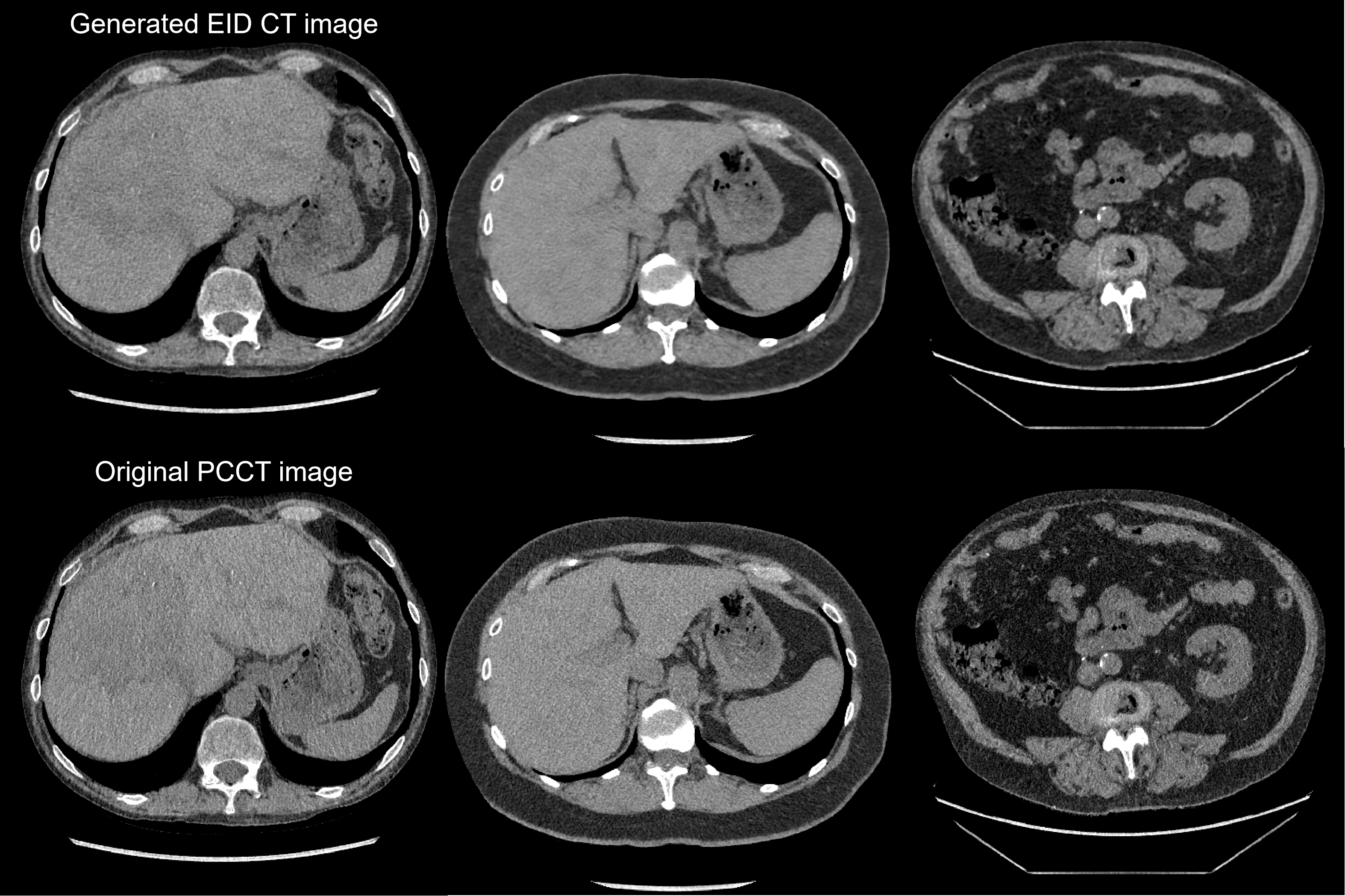}
\end{tabular}
\end{center}
\caption 
{ \label{fig1}
Example of generated EID CT images by Step 1. The images at the first row are generated EID CT images, the images at the second row are original PCCT images. The window/level is 400/40 HU. } 
\end{figure}

The results of Step 1 are presented in Fig.~\ref{fig1}. Compared with the original PCCT image, the synthesized EID CT image appears slightly more blurred, but it still preserves the main anatomical structures and overall tissue textures of the PCCT image.

\begin{figure}[t]
\centering

\includegraphics[height=5.5cm,keepaspectratio]{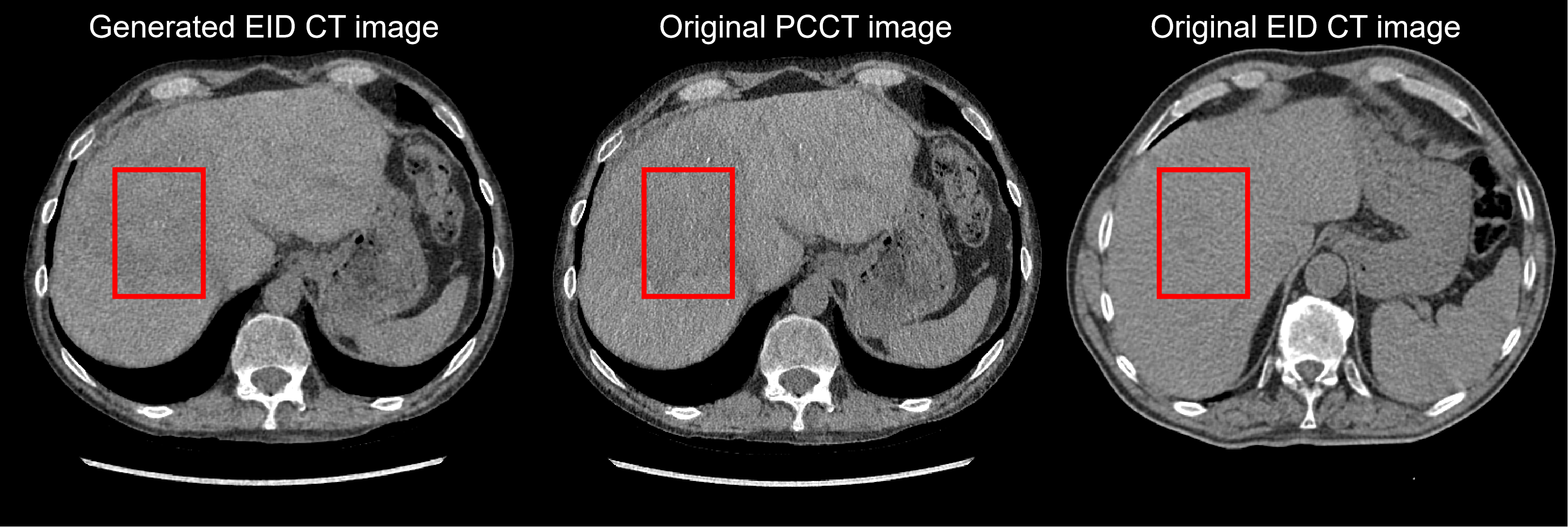}
(a)\\[2mm]

\includegraphics[width=0.9\textwidth,keepaspectratio]{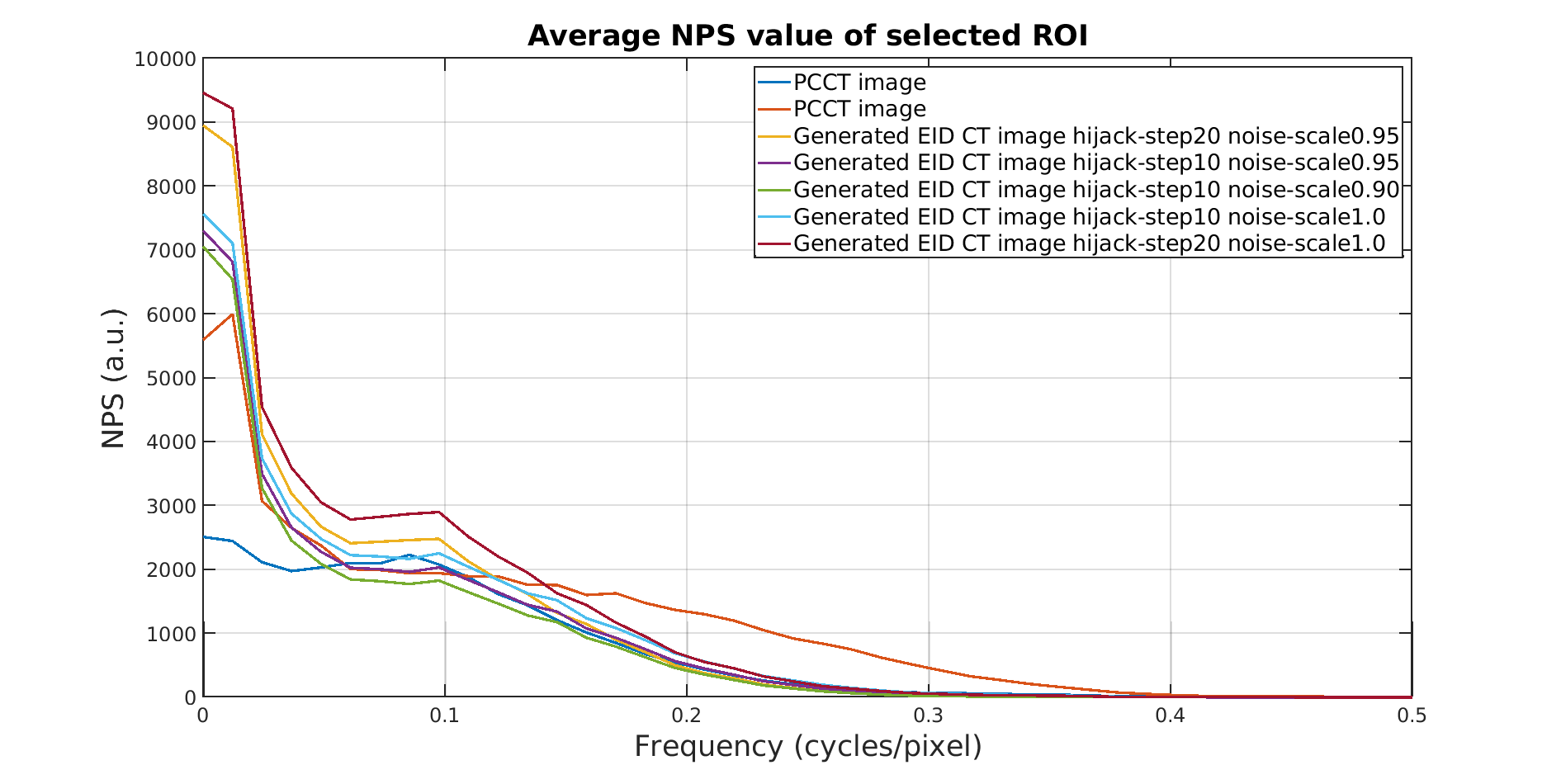}\\
(b)
\caption{Comparison of the mean NPS measured in matched ROIs for EID CT, PCCT, and the generated images. (a) Example slice showing the selected ROI (red rectangle). The generated EID CT image and original PCCT image are paired. The rightmost EID CT image is a slice from a different patient at a comparable anatomical level and is included for reference. (b) Mean NPS curves computed from the ROIs.}
\label{fig2}
\end{figure}

The average NPS results are shown in Fig.~\ref{fig2}. A representative slice and the corresponding ROI selection are illustrated in Fig.~\ref{fig2}(a) . We compared the mean NPS of synthesized EID CT images generated using different step 1 settings, including the hijack timestep (the reverse-time step at which the PCCT image is injected) and the noise-strength factor. For clarity, only representative parameter combinations are shown in the plot. Overall, the parameters setting the hijack step to 10 and the noise scale factor to 0.95 provided a balanced spectrum. It showed closer agreement with the real EID CT curve at higher spatial frequencies, where fine-grain noise texture is expected to contribute more prominently, while maintaining low-frequency behavior closer to the PCCT input.  We therefore used a hijack step of 10 and a noise scale factor of 0.95 to generate the pseudo-paired training images for step 2.

\begin{figure}[!htbp]
\centering
\includegraphics[width=4.5in]{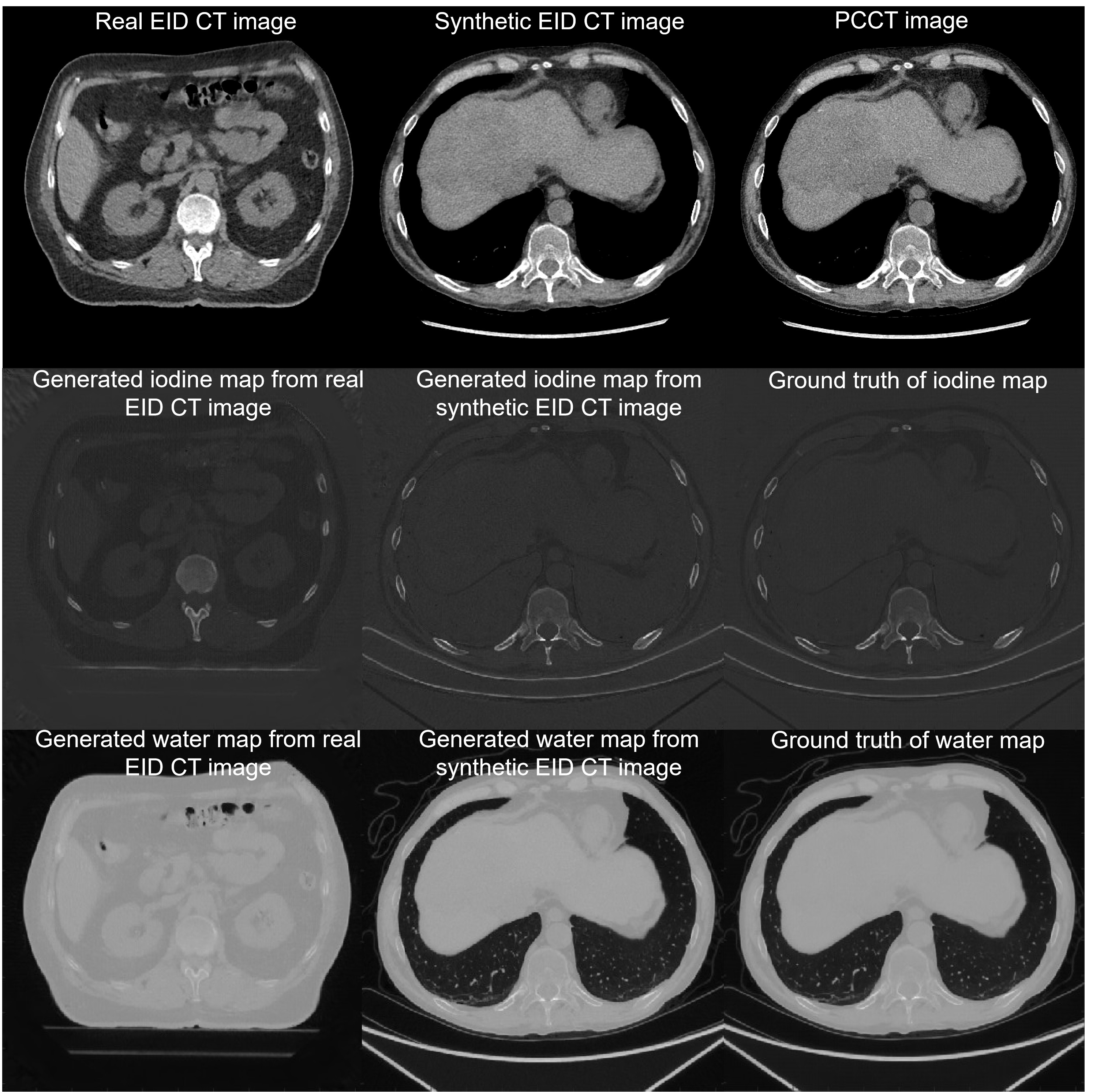}
\caption{Example results from the U-Net model. The left column shows predictions obtained from real EID CT inputs from the test set. The middle column shows predictions obtained from synthetic EID CT inputs from the validation set (Step 1 outputs). The right column shows the corresponding ground truth from the validation set. The window/level of CT images is 400/40 HU. The window/level of iodine map is 300/80.The window/level of water map is 1800/800.}\label{fig3}
\end{figure}

The Step 2 results are presented in Fig.~\ref{fig3}. In rows 2–3, the first column shows the results obtained when the input is a real EID CT image, the second column shows the results obtained when the input is a synthetic EID CT image from the validation set (Step-1 output), and the third column displays the corresponding ground-truth material-basis maps used for training. We observe that the network successfully predicts both the water and iodine maps from the input images, with anatomically consistent patterns and physically plausible value ranges.

\begin{table}[!t]
\centering
\caption{Quantitative results on 70~keV and 40~keV VMIs from the validation set under different training settings.
Higher PSNR/SSIM indicates better fidelity, while lower LPIPS indicates better perceptual similarity.
The loss-term weights for Experiments~1--3 are given as
$(\lambda_{\mathrm{DANN}},\,\lambda_{W},\,\lambda_{I})$:
Exp.~1: $(1,4,40)$; Exp.~2: $(1,2,40)$; Exp.~3: $(1,4,60)$.}
\label{tab:metrics_40_70}

\begin{tabular*}{\linewidth}{@{\extracolsep{\fill}} l c c c c c c @{}}
\toprule
\multirow{2}{*}{Experiment} & \multicolumn{3}{c}{70~keV} & \multicolumn{3}{c}{40~keV} \\
\cmidrule(r){2-4}\cmidrule(l){5-7}
 & PSNR $\uparrow$ & SSIM $\uparrow$ & LPIPS $\downarrow$
 & PSNR $\uparrow$ & SSIM $\uparrow$ & LPIPS $\downarrow$ \\
\midrule
Experiment 1 & \textbf{40.8954} & \textbf{0.9716} & 0.0304 & \textbf{36.4790} & \textbf{0.9209} & 0.0765 \\
Experiment 2 & 38.4468 & 0.9673 & 0.0291 & 35.1640 & 0.9120 & 0.0693 \\
Experiment 3 & 36.4715 & 0.9666 & \textbf{0.0197} & 26.5886 & 0.8814 & \textbf{0.0495} \\
\bottomrule
\end{tabular*}
\end{table}

We computed 40 keV and 70 keV VMIs from the network-predicted material-basis maps on the validation set. Since the validation set provides paired data (synthetic EID CT inputs with corresponding ground-truth spectral material-basis images), we were able to perform full-reference quantitative evaluation using PSNR, SSIM, and LPIPS. PSNR and SSIM measure pixel/structure fidelity (higher is better), whereas LPIPS is a perceptual feature-based distance (lower is better). The quantitative results for different training settings are summarized in Table~\ref{tab:metrics_40_70}. The parameter setting with loss weights $(\lambda_{\mathrm{DANN}},\,\lambda_{W},\,\lambda_{I})=(1,4,40)$ achieved the best PSNR and SSIM, but produced a worse LPIPS value. Qualitative comparisons of the predicted and ground-truth 40 keV and 70 keV VMIs are shown in Fig.~\ref{fig4}(a) , and the corresponding horizontal and vertical mid-line profiles are shown in Fig.~\ref{fig4}(b) and Fig.~\ref{fig4}(c). As shown in Fig.~\ref{fig5}, three representative ROIs were selected on a single example slice for HU bias assessment. The ROI-based mean and standard deviation of HU values for the original and generated VMIs are reported in Table~\ref{tab:roi_hu_bias}. This analysis was performed only on the representative slice shown in Fig.~\ref{fig5}, rather than across multiple images. The predicted 40 and 70 keV VMIs show high consistency with ground truth images in terms of overall attenuation and anatomy structure. However, the predicted images appear slightly over-smoothed compared with the reference, with reduced fine-grained texture and high-frequency detail. This loss of local texture can lead to a worse LPIPS score. The line profiles indicate that the predicted attenuation values follow the ground truth closely, supporting that the reconstructed VMIs are quantitatively reasonable.

Table~\ref{tab:mind_70_eid} reports the MIND-based similarity between the generated 70~keV VMIs and the corresponding real EID CT images. As shown in Table~\ref{tab:mind_70_eid}, the configuration with a loss-weight setting of $(\lambda_{\mathrm{DANN}},\,\lambda_{W},\,\lambda_{I})=(1,4,40)$  achieves the best MIND score, suggesting that it preserves anatomical structures most effectively among the different parameter settings. 

The Step 2 results for real EID CT inputs are shown in Fig.~\ref{fig6}. The examples in Fig.~\ref{fig6}(a) suggest that the network largely preserves the anatomical structures and overall texture of the original EID CT while producing VMIs with improved visual clarity and reduced noise in 70 keV VMI. This result is consistent with prior observations that PCCT can provide lower-noise images and improved noise characteristics compared with conventional EID CT. Moreover, the higher attenuation of bone regions observed at 40 keV compared with 70 keV is physically expected in VMIs. As shown in Fig.~\ref{fig6}(b) and Fig.~\ref{fig6}(c), the horizontal and vertical middle-line profiles of the generated 40 keV and 70 keV VMIs exhibit sharper transitions with higher peaks and deeper valleys than the original EID CT, indicating reasonable network predictions. Figure~\ref{fig8} shows the MTF results obtained from three representative ROIs shown in Fig.~\ref{fig7}. ROI 1 corresponds to the body-boundary edge, as shown in Fig.~\ref{fig8}(a); ROI 2 corresponds to the air--tissue edge, as shown in Fig.~\ref{fig8}(b); and ROI 3 corresponds to the cortical-bone edge, as shown in Fig.~\ref{fig8}(c). For ROI 1, the MTF curves of the original EID CT image and the generated VMI images were highly similar, indicating comparable spatial-frequency preservation at the body boundary. For ROI 2 and ROI 3, the generated 40 and 70 keV VMI images showed higher MTF values than the original EID CT image. This suggests improved apparent edge sharpness and better preservation of high-spatial-frequency information in the generated VMI images, indicating higher apparent spatial resolution.

\begin{figure}[!htbp]
\centering
\includegraphics[width=0.7\linewidth]{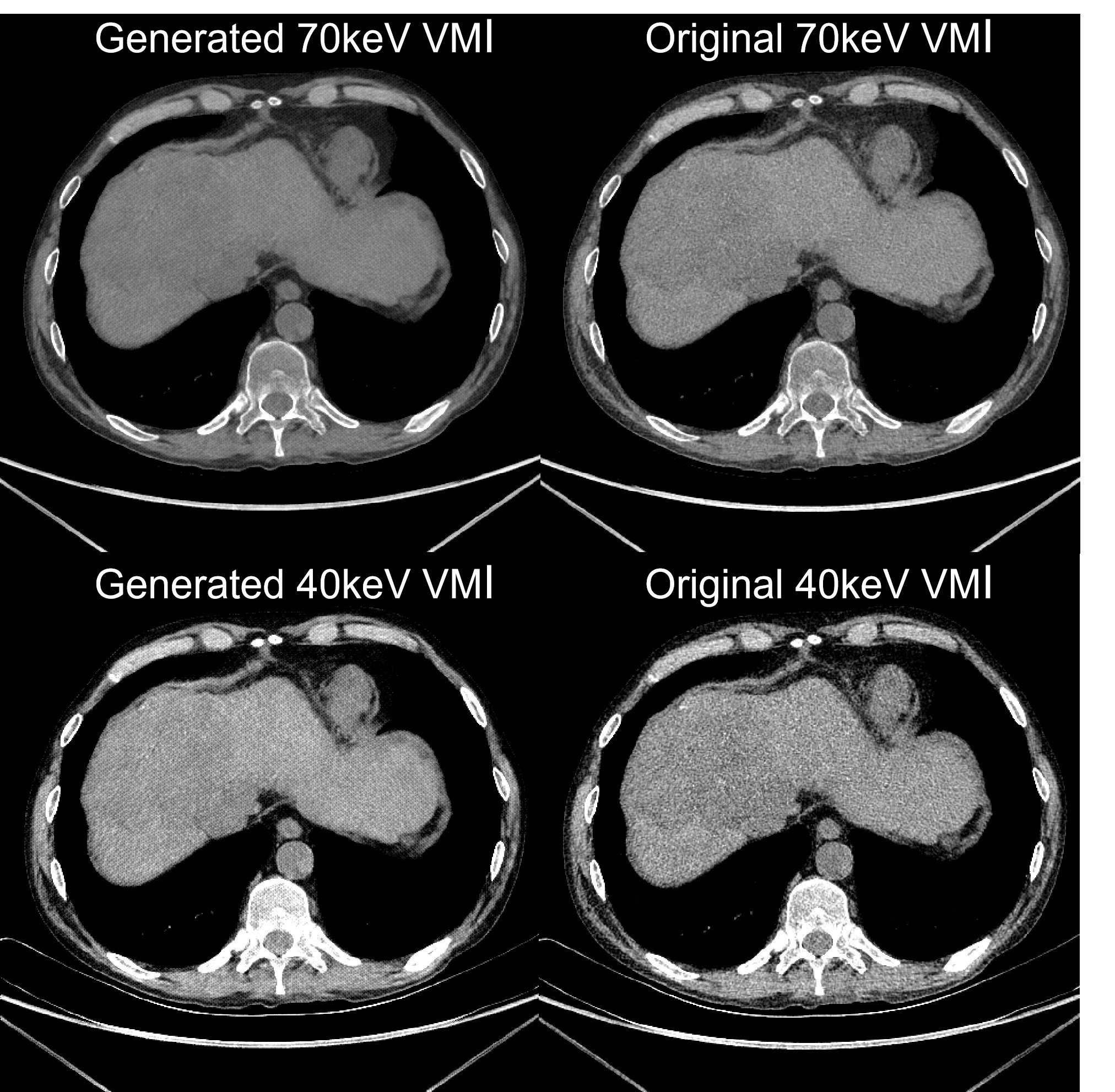}\\
(a)\\[2mm]
\includegraphics[width=0.9\linewidth]{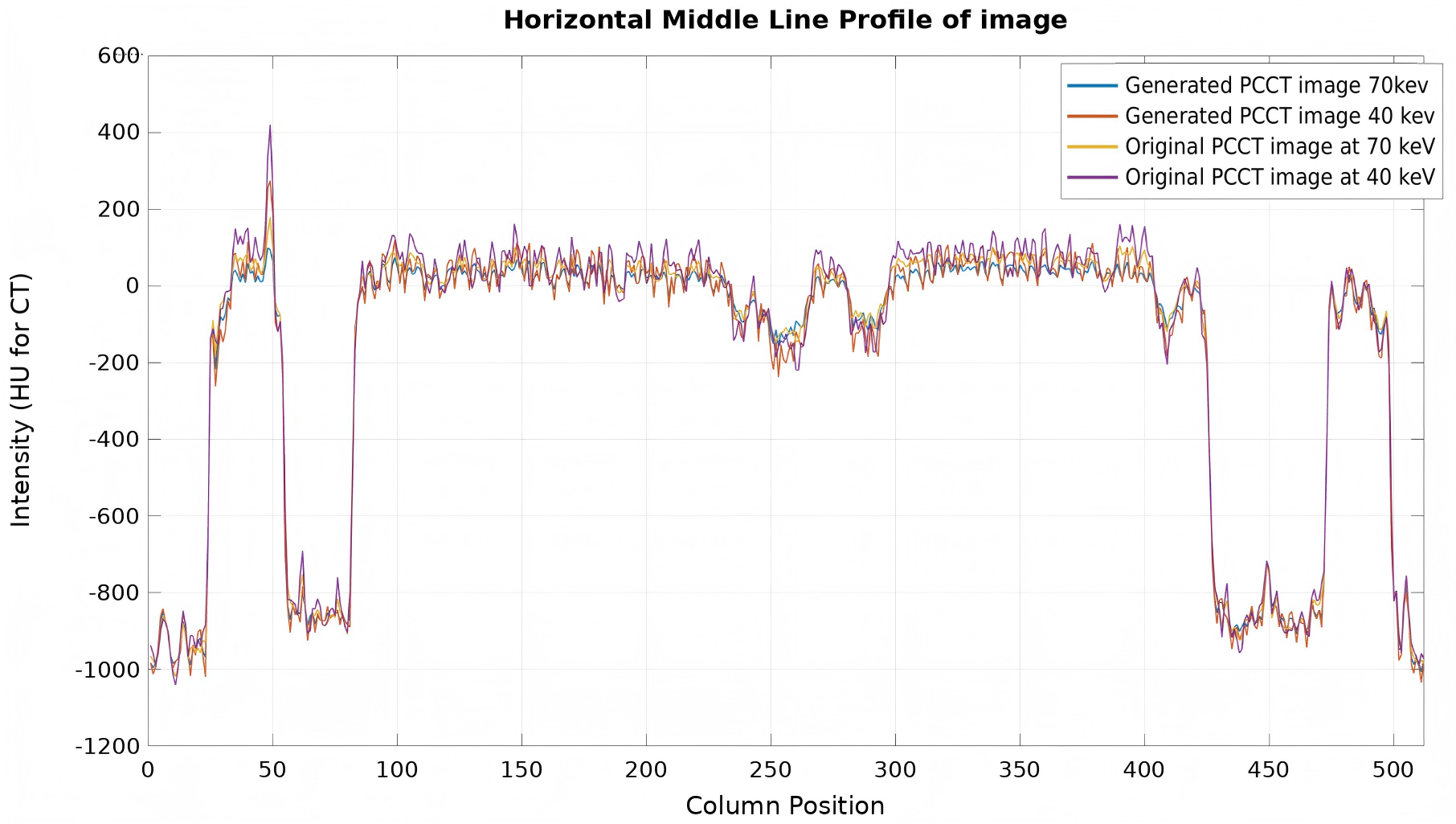}\\
(b)
\end{figure}

\clearpage  

\begin{figure}[t]
\centering
\includegraphics[width=0.85\linewidth]{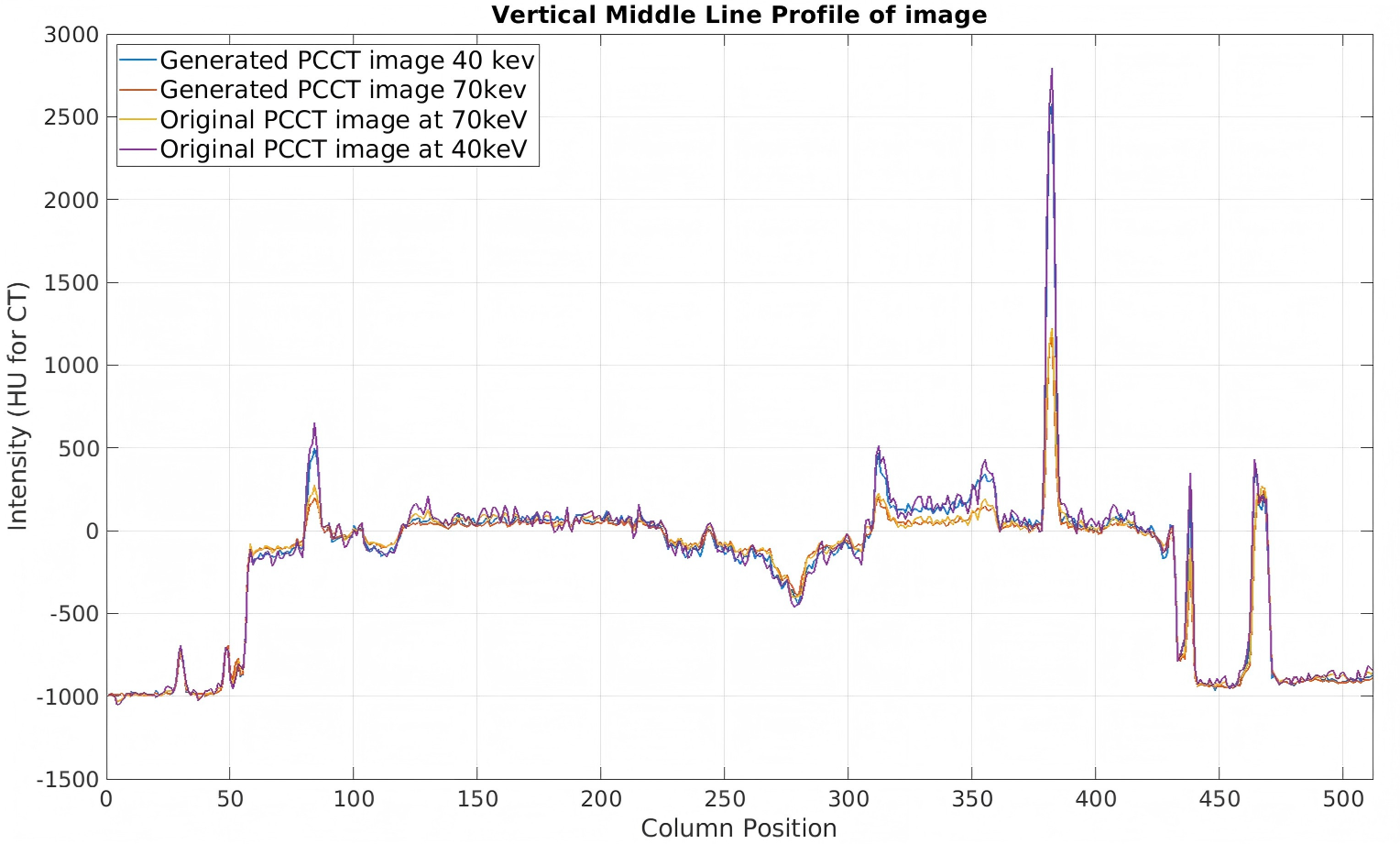}\\
(c)
\caption{Example of 40 and 70 keV VMIs from validation dataset. (a) 40 and 70 keV VMIs of validation dataset and corresponding ground truth. The window/level is 400/40 HU. (b) Horizontal middle line profile of images shown in (a). (c) Vertical middle line profile of images shown in (a).}
\label{fig4}
\end{figure}

\begin{figure}[!htbp]
    \centering
    \includegraphics[width=0.65\textwidth]{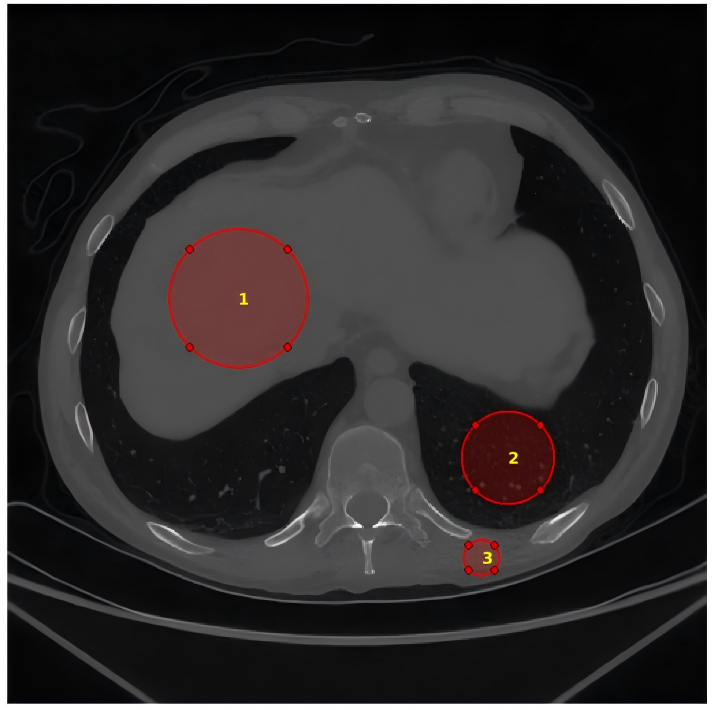}
    \caption{Representative image showing the three ROIs used for HU bias assessment.}
    \label{fig5}
\end{figure}

\begin{table}[!htbp]
\centering
\caption{ROI-based HU bias assessment for original and generated VMI images. ROIs are shown in Fig.~\ref{fig5}. Values are reported as mean $\pm$ standard deviation (HU).}
\label{tab:roi_hu_bias}
\resizebox{\textwidth}{!}{
\begin{tabular}{ccccc}
\hline
\textbf{ROI} & \textbf{Original 70 keV} & \textbf{Generated 70 keV} & \textbf{Original 40 keV} & \textbf{Generated 40 keV} \\
\hline
ROI 1 & $48.2 \pm 21.7$ & $42.0 \pm 18.4$ & $58.4 \pm 42.7$ & $51.8 \pm 32.8$ \\
ROI 2 & $-816.8 \pm 81.6$ & $-823.7 \pm 76.3$ & $-803.4 \pm 86.7$ & $-819.5 \pm 78.3$ \\
ROI 3 & $41.0 \pm 33.6$ & $39.7 \pm 28.9$ & $59.3 \pm 59.8$ & $52.8 \pm 50.3$ \\
\hline
\end{tabular}
}
\end{table}
\begin{table}[!htbp]
\centering
\caption{MIND distance between the predicted 70~keV image from the test dataset and the original EID CT image (lower is better). The loss-term weights for Experiments~1--3 are given as
$(\lambda_{\mathrm{DANN}},\,\lambda_{W},\,\lambda_{I})$:
Exp.~1: $(1,4,40)$; Exp.~2: $(1,2,40)$; Exp.~3: $(1,4,60)$.}
\label{tab:mind_70_eid}

\begin{tabular*}{\linewidth}{@{\extracolsep{\fill}} l c @{}}
\toprule
Experiment & MIND $\downarrow$ \\
\midrule
Experiment 1 & \textbf{1.3280} \\
Experiment 2   & 1.3988 \\
Experiment 3 & 1.3288 \\
\bottomrule
\end{tabular*}
\end{table}

\begin{figure}[!htbp]
\centering
\includegraphics[height=5.5cm,keepaspectratio]{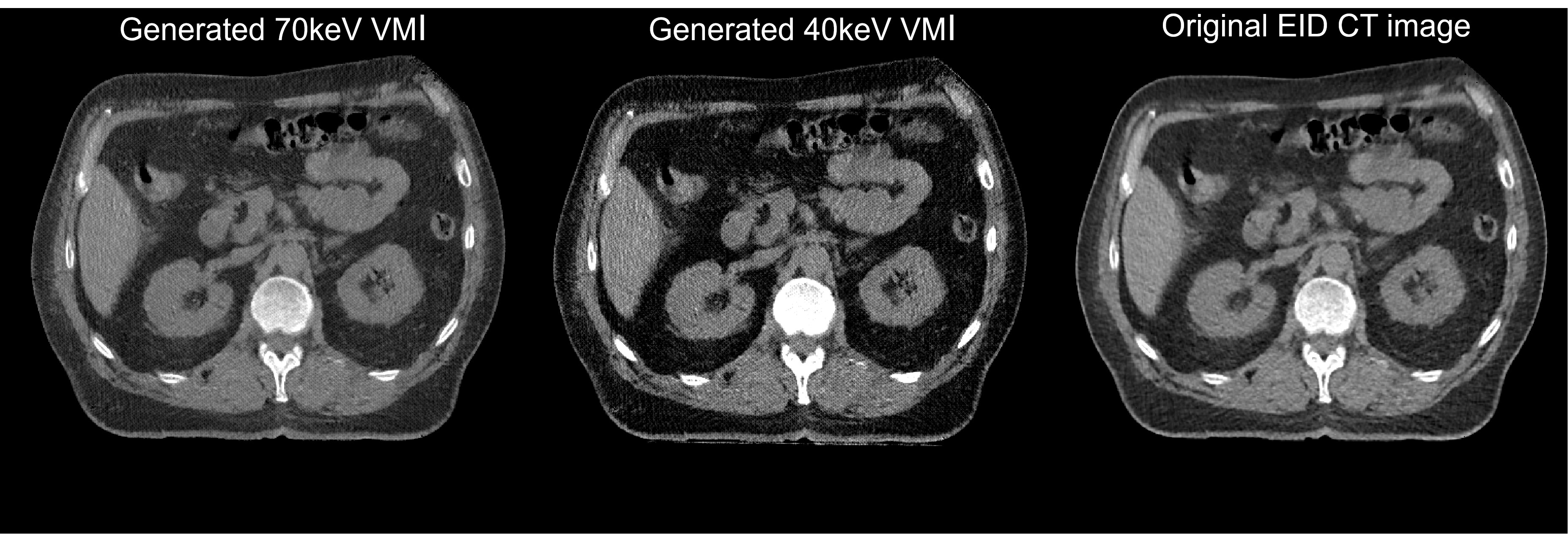}\\
(a)
\phantomcaption
\label{fig6}
\end{figure}

\clearpage

\begin{figure}[!htbp]
\ContinuedFloat
\centering
\includegraphics[width=0.85\textwidth,keepaspectratio]{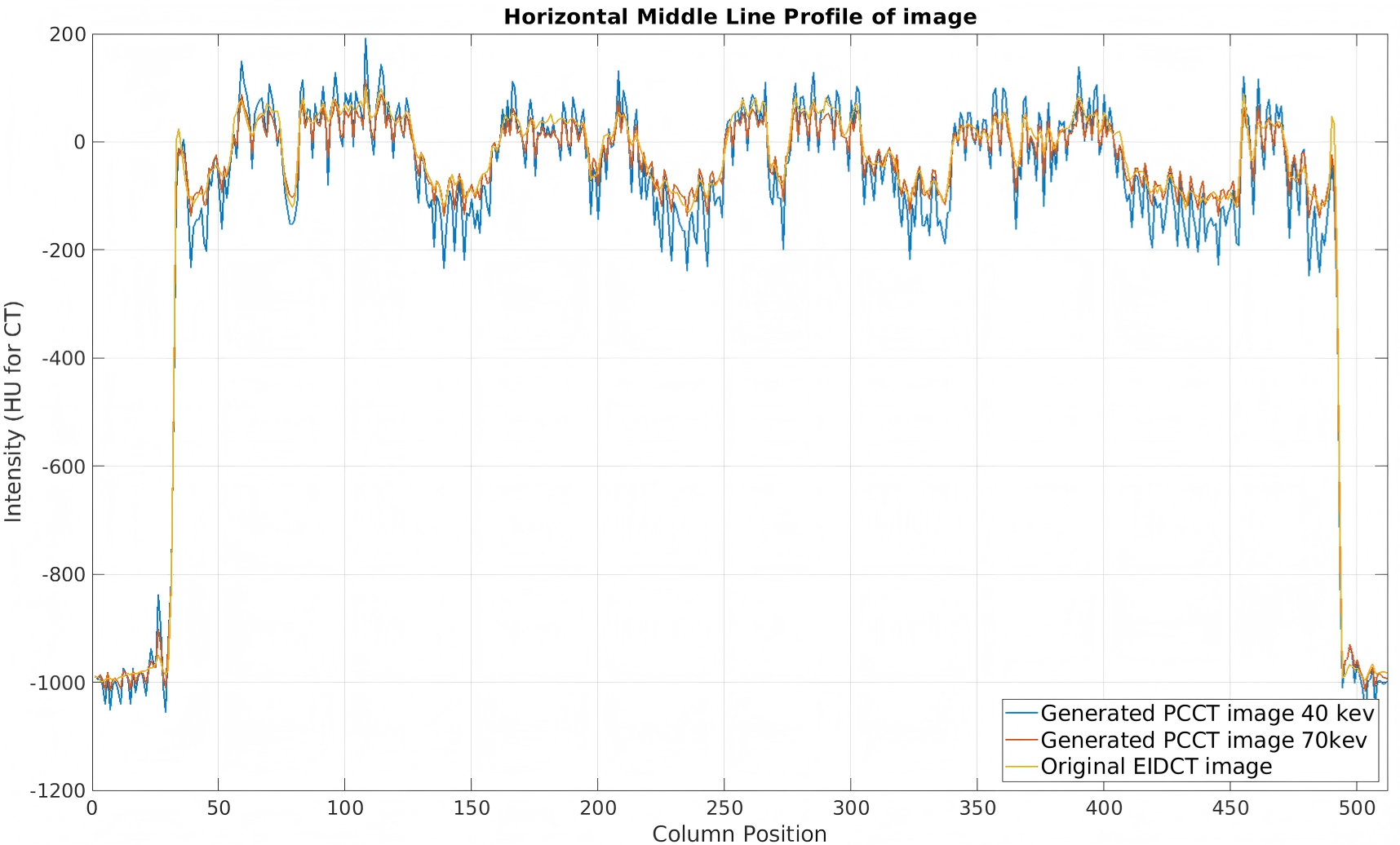}\\
(b)\\[2mm]
\includegraphics[width=0.9\textwidth,keepaspectratio]{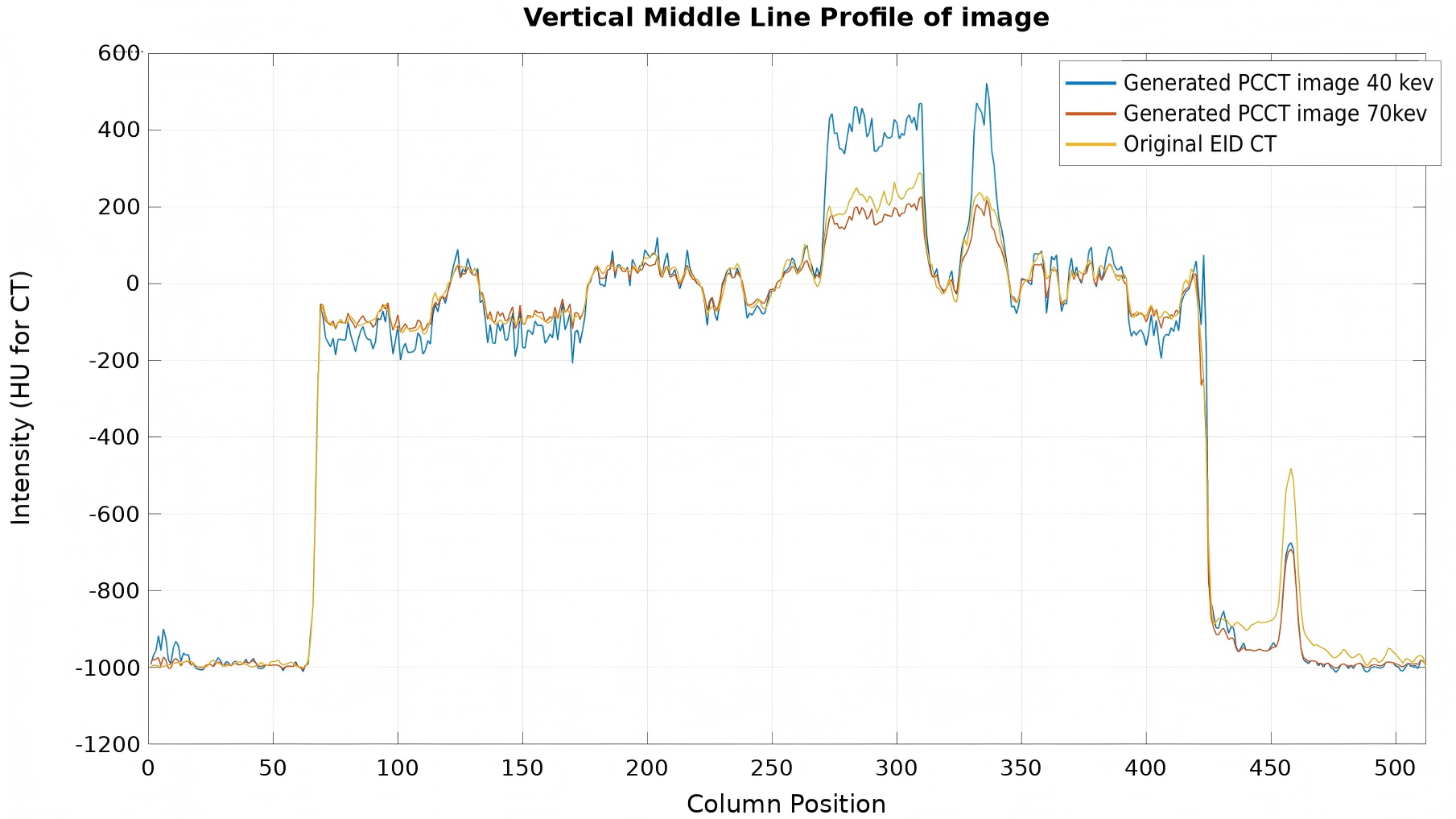}\\
(c)
\caption{Examples of results on real EID CT images. (a) Predicted 40 and 70 keV VMIs and original EID CT image. The window/level is 400/40 HU. (b) Horizontal middle line profile of original EID CT image and generated 40 keV and 70 keV PCCT image. (c) Vertical middle line profile of original EID CT image and generated 40 keV and 70 keV PCCT image.}
\end{figure}

\begin{figure}[!htbp]
    \centering
    \includegraphics[width=0.65\textwidth]{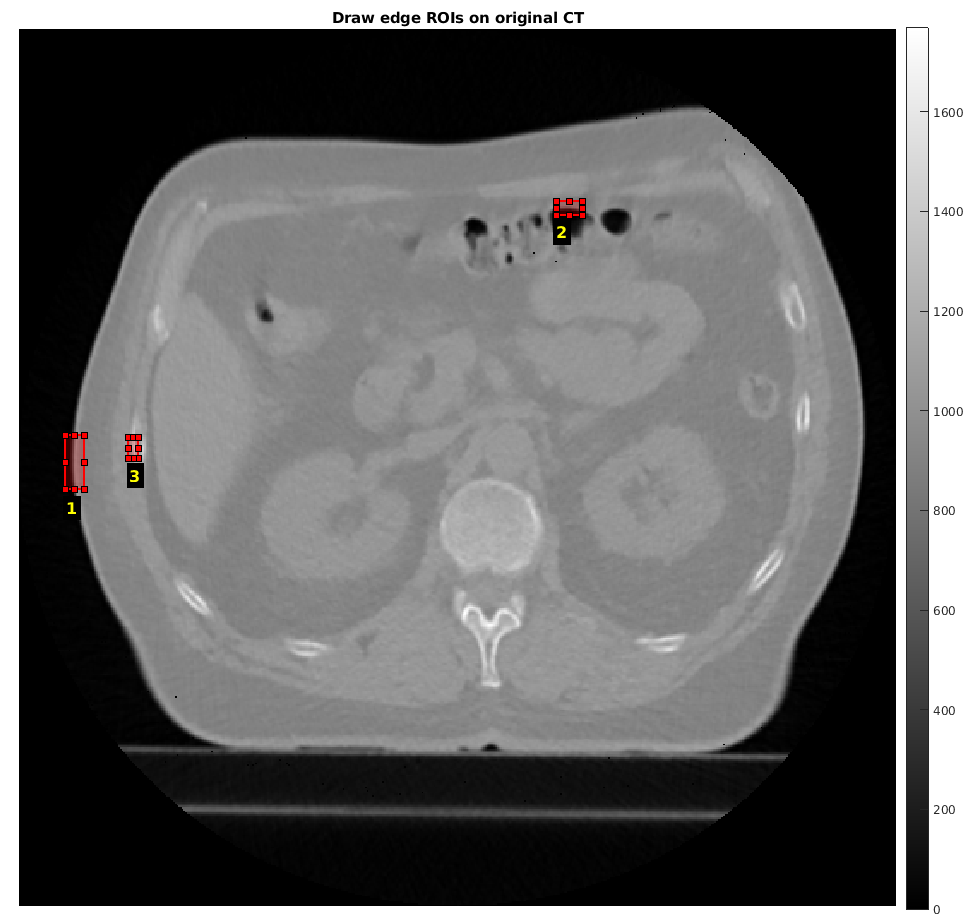}
    \caption{Representative image showing the three ROIs used for HU bias assessment.}
    \label{fig7}
\end{figure}

\begin{figure}[!htbp]
    \centering
    \begin{subfigure}[b]{0.33\textwidth}
        \centering
        \includegraphics[width=\textwidth]{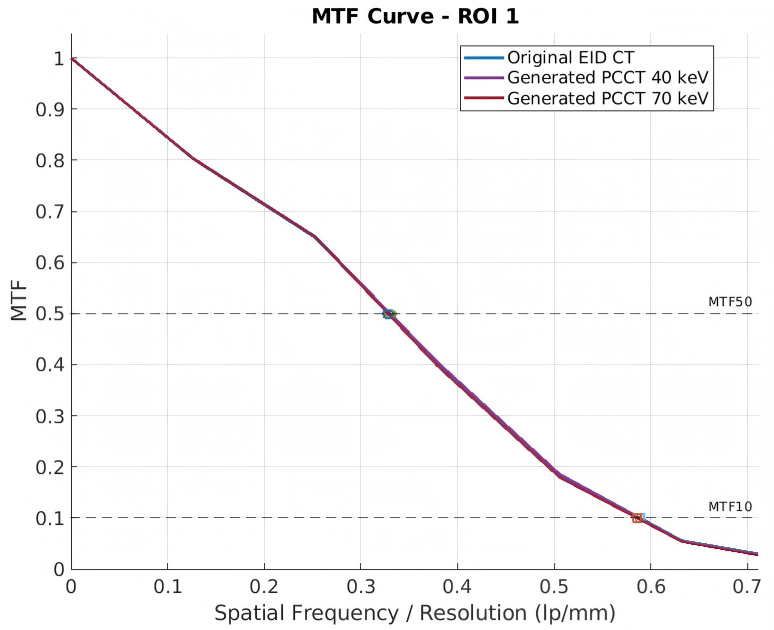}
        \caption{}
        \label{fig:example_a}
    \end{subfigure}
    \hspace{-0.01\textwidth}
    \begin{subfigure}[b]{0.33\textwidth}
        \centering
        \includegraphics[width=\textwidth]{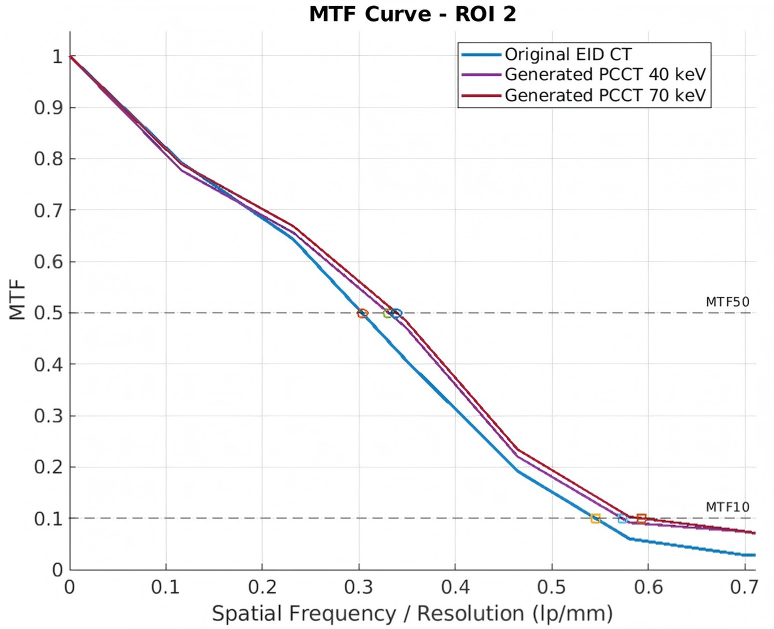}
        \caption{}
        \label{fig:example_b}
    \end{subfigure}
    \hspace{-0.01\textwidth}
    \begin{subfigure}[b]{0.33\textwidth}
        \centering
        \includegraphics[width=\textwidth]{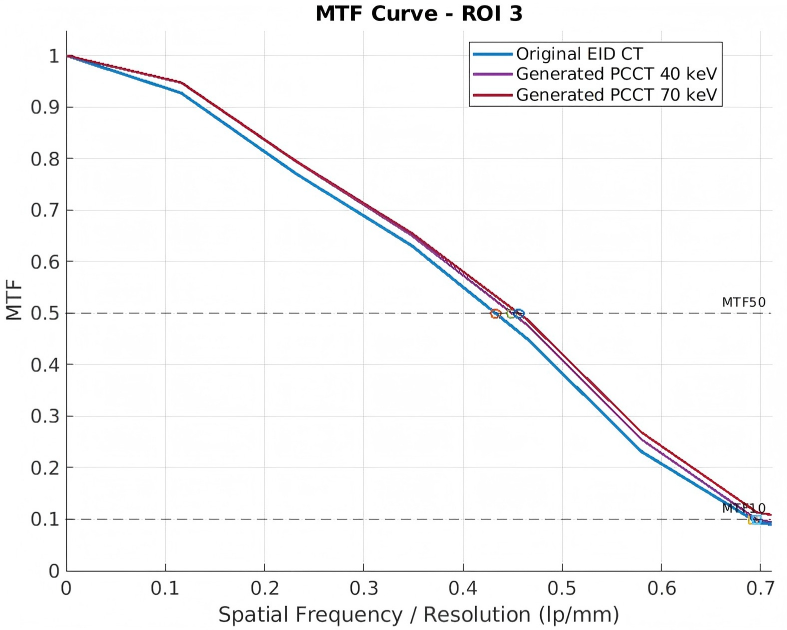}
        \caption{}
        \label{fig:example_c}
    \end{subfigure}
    \caption{MTF analysis based on three representative regions of interest (ROIs). ROIs are shown in Fig.~\ref{fig7}.(a) ROI 1 selected at the body-boundary edge. (b) ROI 2 selected at the air--tissue edge. (c) ROI 3 selected at the cortical-bone edge.}
    \label{fig8}
\end{figure}

\section{Discussion and conclusion}\label{sec12}

Synthesizing clinical images is an important topic in medical imaging, particularly when access to real data is limited. However, PCCT image synthesis from conventional EID CT remains underexplored. In this work, we propose a two-step framework to synthesize spectral material-basis images from conventional EID CT images, and our experiments demonstrate the feasibility of the proposed approach. Overall, the model preserves major anatomical structures while shifting the image appearance—particularly noise texture and perceived resolution—toward a PCCT-like style. A practical advantage of the proposed framework is that it can be trained with only a small PCCT dataset, which is particularly useful when prototyping or evaluating new image modes where only limited data are initially available.

A limitation of our work is that the synthesized images tend to exhibit unnatural texture, mild artifacts, and over-smoothing. We hypothesize that these problems stem from two main reasons. First, the choice and weighting of the loss terms may favor pixel-wise fidelity and structural similarity, which can suppress high-frequency details and lead to overly smooth outputs; perceptual metrics such as LPIPS are known to be sensitive to these texture differences. Second, the synthetic EID CT images produced in Step 1 may still differ from the real EID CT distribution. This is because the diffusion model is trained only on EID CT images and does not observe PCCT images during training. We then injected a PCCT image into the reverse DDIM sampling trajectory to obtain an EID-like output, which can leave a residual domain gap and make it more difficult for the Step-2 network to learn truly domain-invariant features.

In future work, we plan to expand the dataset and further improve the method to enhance image fidelity and robustness. One practical next step is to leverage the larger pool of available EID CT images to train the Step 1 diffusion model on more diverse anatomy and acquisition conditions, which may improve robustness and generalization. In addition, we will explore alternative unpaired learning strategies for mapping PCCT to the EID CT domain, such as CycleGAN-based translation and related unsupervised domain adaptation approaches. 

While we only focus on non-enhanced chest and abdomen imaging in the current study, the next step is to extend the framework to contrast enhanced CT and to other body parts. This larger anatomical variability  may require additional training data and domain adaptation strategies.

In conclusion, we have demonstrated a feasible approach for synthesizing PCCT material-basis images and VMIs from conventional EID CT, requiring only a small PCCT dataset together with an unpaired EID CT dataset for training. This is particularly useful for developing and adapting deep-learning methods for PCCT applications, including situations where new imaging modes or protocols are being explored and only limited PCCT data are initially available.

\subsection*{Disclosures}
Mats Persson and Ruihan Huang disclose research collaboration with GE Health-Care. Mats Persson discloses research support and license fees, GE HealthCare.

\subsection* {Code, Data, and Materials Availability} 
The clinical data were provided by Karolinska University Hospital and GE HealthCare. The data that support the findings of this study are not publicly available due to institutional and contractual restrictions. Code and data are available under conditions of confidentiality, contingent on approval by GE HealthCare, and can be requested from the author at ruihanh@ug.kth.se.

\subsection* {Acknowledgments}
The study was approved by the Swedish Ethics Review Authority (permits 2023-03709-01, 2024-02038-02, and 2024-07310-02) and is registered at www.clinicaltrials.gov with NCT registration number NCT05835284. Mats Persson and Ruihan Huang disclose research collaboration with GE HealthCare. Mats Persson discloses research support and license fees, GE HealthCare. This study received financial support from MedTechLabs, The Swedish Research Council (grant 2021-05103), the China Scholarship Council, and the Göran Gustafsson Foundation (grants 2114, 2210 and 2307). The authors
also would like to thank Hugo Rippe, Edvard Oxelström and Axel Riley for helpful discussions.



\bibliography{report}

@article{Almqvist2024SecondGenSiPCCT,
  author  = {Almqvist, Hakan and Crotty, Dominic and Nyren, Sven and Yu, Jimmy and Arnberg-Sandor, Fabian and Brismar, Torkel and Tovatt, Cedric and Linder, Hugo and Dagotto, Jose and Fredenberg, Erik and Tamm, Moa Yveborg and Deak, Paul and Fanariotis, Michail and Bujila, Robert and Holmin, Staffan},
  title   = {Initial Clinical Images From a Second-Generation Prototype Silicon-Based Photon-Counting Computed Tomography System},
  journal = {Academic Radiology},
  year    = {2024},
  month   = feb,
  volume  = {31},
  number  = {2},
  pages   = {572--581},
  doi     = {10.1016/j.acra.2023.06.031},
  pmid    = {37563023},
  note    = {Epub 2023-08-08}
}

@article{Danielsson2021PhotonCountingDetectorsCT,
  author  = {Danielsson, Mats and Persson, Mats and Sj{\"o}lin, Martin},
  title   = {Photon-counting x-ray detectors for CT},
  journal = {Physics in Medicine and Biology},
  year    = {2021},
  month   = jan,
  volume  = {66},
  number  = {3},
  pages   = {03TR01},
  doi     = {10.1088/1361-6560/abc5a5},
  pmid    = {33113525}
}

@article{Johnstone2018SystematicReviewSyntheticCT,
  author  = {Johnstone, Emily and Wyatt, Jonathan J. and Henry, Ann M. and Short, Susan C. and Sebag-Montefiore, David and Murray, Louise and Kelly, Charles G. and McCallum, Hazel M. and Speight, Richard},
  title   = {Systematic Review of Synthetic Computed Tomography Generation Methodologies for Use in Magnetic Resonance Imaging-Only Radiation Therapy},
  journal = {International Journal of Radiation Oncology, Biology, Physics},
  year    = {2018},
  month   = jan,
  volume  = {100},
  number  = {1},
  pages   = {199--217},
  doi     = {10.1016/j.ijrobp.2017.08.043},
  pmid    = {29254773},
  note    = {Epub 2017-09-08}
}

@article{Yi2019GANMedicalImagingReview,
  author  = {Yi, Xin and Walia, Ekta and Babyn, Paul},
  title   = {Generative adversarial network in medical imaging: A review},
  journal = {Medical Image Analysis},
  year    = {2019},
  month   = dec,
  volume  = {58},
  pages   = {101552},
  doi     = {10.1016/j.media.2019.101552},
  pmid    = {31521965},
  note    = {Epub 2019-08-31.  }
}

@incollection{Nie2017ContextAwareGAN,
  author    = {Nie, Dong and Trullo, Roger and Lian, Jun and Petitjean, Caroline and Ruan, Su and Wang, Qian and Shen, Dinggang},
  title     = {Medical Image Synthesis with Context-Aware Generative Adversarial Networks},
  booktitle = {Medical Image Computing and Computer Assisted Intervention -- MICCAI 2017},
  editor    = {Descoteaux, Maxime and Maier-Hein, Lena and Franz, Alfred and Jannin, Pierre and Collins, D. Louis and Duchesne, Simon},
  series    = {Lecture Notes in Computer Science},
  volume    = {10435},
  pages     = {417--425},
  year      = {2017},
  publisher = {Springer},
  address   = {Cham},
  doi       = {10.1007/978-3-319-66179-7\_48}
}

@incollection{Wolterink2017UnpairedMRtoCT,
  author    = {Wolterink, Jelmer M. and Dinkla, Anna M. and Savenije, Mark H. F. and Seevinck, Peter R. and van den Berg, Cornelis A. T. and I{\v{s}}gum, Ivana},
  title     = {Deep MR to CT Synthesis Using Unpaired Data},
  booktitle = {Simulation and Synthesis in Medical Imaging},
  editor    = {Tsaftaris, Sotirios A. and Gooya, Ali and Frangi, Alejandro F. and Prince, Jerry L.},
  series    = {Lecture Notes in Computer Science},
  volume    = {10557},
  pages     = {14--23},
  year      = {2017},
  publisher = {Springer},
  address   = {Cham},
  doi       = {10.1007/978-3-319-68127-6\_2}
}

@article{Pan2024TransformerDiffusionSyntheticCT,
  author  = {Pan, Shaoyan and Abouei, Elham and Wynne, Jacob and Chang, Chih-Wei and Wang, Tonghe and Qiu, Richard L. J. and Li, Yuheng and Peng, Junbo and Roper, Justin and Patel, Pretesh and Yu, David S. and Mao, Hui and Yang, Xiaofeng},
  title   = {Synthetic CT generation from MRI using 3D transformer-based denoising diffusion model},
  journal = {Medical Physics},
  year    = {2024},
  month   = apr,
  volume  = {51},
  number  = {4},
  pages   = {2538--2548},
  doi     = {10.1002/mp.16847},
  pmid    = {38011588},
  pmcid   = {PMC10994752},
  note    = {Epub 2023-11-27}
}

@article{Li2023QualityCheckedPhysicsConstrained,
  author  = {Li, Yinsheng and Tie, Xin and Li, Ke and Zhang, Ran and Qi, Zhihua and Budde, Adam and Grist, Thomas M. and Chen, Guang-Hong},
  title   = {A quality-checked and physics-constrained deep learning method to estimate material basis images from single-kV contrast-enhanced chest CT scans},
  journal = {Medical Physics},
  year    = {2023},
  volume  = {50},
  number  = {6},
  pages   = {3368--3388},
  doi     = {10.1002/mp.16352}
}

@article{Lyu2021EstimatingDualEnergyCT,
  author  = {Lyu, Tianling and Zhao, Wei and Zhu, Yinsu and Wu, Zhan and Zhang, Yikun and Chen, Yang and Luo, Limin and Li, Shuo and Xing, Lei},
  title   = {Estimating dual-energy CT imaging from single-energy CT data with material decomposition convolutional neural network},
  journal = {Medical Image Analysis},
  year    = {2021},
  volume  = {70},
  pages   = {102001},
  doi     = {10.1016/j.media.2021.102001}
}

@article{Li2026ImageQualityAssessmentDLVME,
  author  = {Li, Ke and Nagpal, Prashant and Mullan, Brian F. and Wu, Yijing and Garrett, John W. and Zhang, Ran and Qi, Zhihua and Chen, Guang-Hong and Grist, Thomas M.},
  title   = {Image Quality Assessment of Deep Learning-Based Virtual Monoenergetic Images From Single-Energy CT Pulmonary Angiography},
  journal = {Journal of Computer Assisted Tomography},
  year    = {2026},
  volume  = {50},
  number  = {2},
  pages   = {263--271},
  doi     = {10.1097/RCT.0000000000001812}
}

@article{Lyu2022CTMRIDiffusion,
  author        = {Lyu, Qing and Wang, Ge},
  title         = {Conversion Between CT and MRI Images Using Diffusion and Score-Matching Models},
  journal       = {arXiv preprint arXiv:2209.12104},
  year          = {2022},
  eprint        = {2209.12104},
  archivePrefix = {arXiv},
  primaryClass  = {eess.IV},
  doi           = {10.48550/arXiv.2209.12104}
}

@misc{Yang2017LCTSC,
  author       = {Yang, Jinzhong and Sharp, Gregory and Veeraraghavan, Harini and Van Elmpt, Wouter and Dekker, Andr{\'e} and Lustberg, Tim and Gooding, Mike},
  title        = {Data from Lung CT Segmentation Challenge 2017 (LCTSC)},
  howpublished = {The Cancer Imaging Archive (TCIA) Dataset},
  year         = {2017},
  doi          = {10.7937/K9/TCIA.2017.3R3FVZ08},
  url          = {https://www.cancerimagingarchive.net/collection/lctsc/},
  note         = {Accessed May 22, 2026}
}

@inproceedings{Paszke2019PyTorch,
  author    = {Paszke, Adam and Gross, Sam and Massa, Francisco and Lerer, Adam and Bradbury, James and Chanan, Gregory and Killeen, Trevor and Lin, Zeming and Gimelshein, Natalia and Antiga, Luca and Desmaison, Alban and K{\"o}pf, Andreas and Yang, Edward and DeVito, Zachary and Raison, Martin and Tejani, Alykhan and Chilamkurthy, Sasank and Steiner, Benoit and Fang, Lu and Bai, Junjie and Chintala, Soumith},
  booktitle = {Advances in Neural Information Processing Systems},
  title     = {PyTorch: An Imperative Style, High-Performance Deep Learning Library},
  
  volume    = {32},
  pages     = {8024--8035},
  year      = {2019},
  doi       = {10.48550/arXiv.1912.01703}
}

@inproceedings{kingma2015adam,
  author    = {Kingma, Diederik P. and Ba, Jimmy},
  booktitle = {International Conference on Learning Representations (ICLR)},
  title     = {Adam: A Method for Stochastic Optimization},
  
  year      = {2015},
  url       = {https://arxiv.org/abs/1412.6980},
  doi       = {10.48550/arXiv.1412.6980}
}

@article{Almqvist2024InitialClinicalImagesSiPCCT,
  author  = {Almqvist, Hakan and Crotty, Dominic and Nyren, Sven and Yu, Jimmy and Arnberg-Sandor, Fabian and Brismar, Torkel and Tovatt, Cedric and Linder, Hugo and Dagotto, Jose and Fredenberg, Erik and Tamm, Moa Yveborg and Deak, Paul and Fanariotis, Michail and Bujila, Robert and Holmin, Staffan},
  title   = {Initial Clinical Images From a Second-Generation Prototype Silicon-Based Photon-Counting Computed Tomography System},
  journal = {Academic Radiology},
  year    = {2024},
  month   = feb,
  volume  = {31},
  number  = {2},
  pages   = {572--581},
  doi     = {10.1016/j.acra.2023.06.031},
  pmid    = {37563023},
  note    = {Epub 2023-08-08},
  url     = {https://pubmed.ncbi.nlm.nih.gov/37563023/}
}

@misc{song2022denoisingdiffusionimplicitmodels,
      title={Denoising Diffusion Implicit Models}, 
      author={Jiaming Song and Chenlin Meng and Stefano Ermon},
      year={2022},
      eprint={2010.02502},
      archivePrefix={arXiv},
      primaryClass={cs.LG},
      howpublished={https://arxiv.org/abs/2010.02502}, 
      doi= {10.48550/arXiv.2010.02502}
}

@inproceedings{ho2020ddpm,
 author = {Ho, Jonathan and Jain, Ajay and Abbeel, Pieter},
 booktitle = {Advances in Neural Information Processing Systems},
 editor = {H. Larochelle and M. Ranzato and R. Hadsell and M.F. Balcan and H. Lin},
 pages = {6840--6851},
 publisher = {Curran Associates, Inc.},
 title = {Denoising Diffusion Probabilistic Models},
 volume = {33},
 year = {2020},
 doi  = {10.48550/arXiv.2006.11239}
}

@inproceedings{ronneberger2015unetconvolutionalnetworksbiomedical,
  author    = {Ronneberger, Olaf and Fischer, Philipp and Brox, Thomas},
  title     = {{U-Net: Convolutional Networks for Biomedical Image Segmentation}},
  booktitle = {{Medical Image Computing and Computer-Assisted Intervention -- MICCAI 2015}},
  series    = {Lecture Notes in Computer Science},
  volume    = {9351},
  pages     = {234--241},
  year      = {2015},
  publisher = {Springer},
  doi       = {10.1007/978-3-319-24574-4\_28}
}

@article{ganin2016domain,
  title={Domain-adversarial training of neural networks},
  author={Ganin, Yaroslav and Ustinova, Evgeniya and Ajakan, Hana and Germain, Pascal and Larochelle, Hugo and Laviolette, Fran{\c{c}}ois and March, Mario and Lempitsky, Victor},
  journal={Journal of machine learning research},
  volume={17},
  number={59},
  pages={1--35},
  year={2016},
  doi= {10.48550/arXiv.1505.07818}
}

@article{Riederer1978NPS,
  author  = {Riederer, Stephen J. and Pelc, Norbert J. and Chesler, David A.},
  title   = {The Noise Power Spectrum in Computed X-Ray Tomography},
  journal = {Physics in Medicine and Biology},
  year    = {1978},
  volume  = {23},
  number  = {3},
  pages   = {446--454},
  doi     = {10.1088/0031-9155/23/3/008}
}

@article{Liu2026SUMI,
  author        = {Liu, Junqi and Zhou, Xinze and Li, Wenxuan and Ye, Scott and Sitek, Arkadiusz and Yang, Xiaofeng and Tang, Yucheng and Xu, Daguang and Ding, Kai and Wang, Kang and Yang, Yang and Yuille, Alan L. and Zhou, Zongwei},
  title         = {Distilling Photon-Counting CT into Routine Chest CT through Clinically Validated Degradation Modeling},
  journal       = {arXiv preprint arXiv:2604.07329},
  year          = {2026},
  eprint        = {2604.07329},
  archivePrefix = {arXiv},
  primaryClass  = {cs.CV},
  doi           = {10.48550/arXiv.2604.07329},
  url           = {https://arxiv.org/abs/2604.07329}
}

@misc{matlab_r2023b,
  author       = {{The MathWorks Inc.}},
  title        = {{MATLAB version R2023b}},
  year         = {2023},
  howpublished = {\url{https://www.mathworks.com/products/matlab.html}},
  note         = {Natick, Massachusetts, United States}
}

@article{wang2004ssim,
  author    = {Wang, Zhou and Bovik, Alan C. and Sheikh, Hamid R. and Simoncelli, Eero P.},
  title     = {Image Quality Assessment: From Error Visibility to Structural Similarity},
  journal   = {IEEE Transactions on Image Processing},
  volume    = {13},
  number    = {4},
  pages     = {600--612},
  year      = {2004},
  doi       = {10.1109/TIP.2003.819861}
}

@inproceedings{zhang2018lpips,
  author    = {Zhang, Richard and Isola, Phillip and Efros, Alexei A. and Shechtman, Eli and Wang, Oliver},
  title     = {The Unreasonable Effectiveness of Deep Features as a Perceptual Metric},
  booktitle = {Proceedings of the IEEE/CVF Conference on Computer Vision and Pattern Recognition},
  pages     = {586--595},
  year      = {2018},
  doi       = {10.1109/CVPR.2018.00068},
  url       = {https://openaccess.thecvf.com/content_cvpr_2018/html/Zhang_The_Unreasonable_Effectiveness_CVPR_2018_paper.html}
}

@article{huynhthu2008psnr,
  author    = {Huynh-Thu, Quan and Ghanbari, Mohammed},
  title     = {Scope of validity of PSNR in image/video quality assessment},
  journal   = {Electronics Letters},
  volume    = {44},
  number    = {13},
  pages     = {800--801},
  year      = {2008},
  doi       = {10.1049/el:20080522}
}

@article{Samei1998PresampledMTF,
  author  = {Samei, Ehsan and Flynn, Michael J. and Reimann, David A.},
  title   = {A method for measuring the presampled MTF of digital radiographic systems using an edge test device},
  journal = {Medical Physics},
  year    = {1998},
  volume  = {25},
  number  = {1},
  pages   = {102--113},
  doi     = {10.1118/1.598165}
}

@article{huang2026pcct_motion_correction,
  author  = {Huang, R. and Larsson, K. and Hein, D. and Holmin, S. and Holmes, T. W. and Pourmorteza, A. and Persson, M.},
  title   = {Deep-learning-based spectral motion artifact correction on photon-counting cardiac CT images},
  journal = {Physics in Medicine \& Biology},
  year    = {2026},
  month   = jan,
  day     = {28},
  doi     = {10.1088/1361-6560/ae3eee},
  note    = {Epub ahead of print. PMID: 41604762}
}

@article{Xu2024WaveletLossCycleGANsVMI,
  author  = {Xu, Zilong and Li, Miaomiao and Li, Baosheng and Shu, Huazhong},
  title   = {Synthesis of virtual monoenergetic images from kilovoltage peak images using wavelet loss enhanced CycleGAN for improving radiomics features reproducibility},
  journal = {Quantitative Imaging in Medicine and Surgery},
  year    = {2024},
  volume  = {14},
  number  = {3},
  pages   = {2370--2390},
  doi     = {10.21037/qims-23-922},
  url     = {https://qims.amegroups.org/article/view/122311/html}
}

@article{Cong2020VirtualMonoenergeticCTPatterns,
  author  = {Cong, Wenxiang and Xi, Yan and Fitzgerald, Paul and De Man, Bruno and Wang, Ge},
  title   = {Virtual Monoenergetic CT Imaging via Deep Learning},
  journal = {Patterns},
  year    = {2020},
  month   = oct,
  volume  = {1},
  number  = {8},
  pages   = {100128},
  doi     = {10.1016/j.patter.2020.100128},
  pmid    = {33294869},
  pmcid   = {PMC7691386},
  url     = {https://www.sciencedirect.com/science/article/pii/S2666389920301690}
}

@article{Kawahara2021MonoenergeticCTGAN,
  author  = {Kawahara, Daisuke and Ozawa, Shuichi and Kimura, Tomoki and Nagata, Yasushi},
  title   = {Image synthesis of monoenergetic CT image in dual-energy CT using kilovoltage CT with deep convolutional generative adversarial networks},
  journal = {Journal of Applied Clinical Medical Physics},
  year    = {2021},
  month   = apr,
  volume  = {22},
  number  = {4},
  pages   = {184--192},
  doi     = {10.1002/acm2.13190},
  pmid    = {33599386},
  pmcid   = {PMC8035569},
  note    = {Epub 2021-02-18},
  url     = {https://pubmed.ncbi.nlm.nih.gov/33599386/}
}

@misc{MathWorks_imregtform_doc,
  author       = {{The MathWorks, Inc.}},
  title        = {\texttt{imregtform}: Estimate geometric transformation that aligns two 2-D or 3-D images},
  howpublished = {MATLAB Help Center, Image Processing Toolbox documentation page},
  year         = {2026},
  url          = {https://www.mathworks.com/help/images/ref/imregtform.html},
  note         = {Accessed February 27, 2026}
}

@inproceedings{Simonyan2015VGG,
  author       = {Simonyan, Karen and Zisserman, Andrew},
  title        = {Very Deep Convolutional Networks for Large-Scale Image Recognition},
  booktitle    = {International Conference on Learning Representations},
  organization = {International Conference on Learning Representations},
  year         = {2015},
  url          = {https://www.robots.ox.ac.uk/~vgg/publications/2015/Simonyan15/},
  doi          = {10.48550/arXiv.1409.1556}
}

@inproceedings{Deng2009ImageNet,
  author    = {Deng, Jia and Dong, Wei and Socher, Richard and Li, Li{-}Jia and Li, Kai and Fei{-}Fei, Li},
  title     = {ImageNet: {A} large-scale hierarchical image database},
  booktitle = {2009 {IEEE} Computer Society Conference on Computer Vision and Pattern Recognition (CVPR 2009), 20--25 June 2009, Miami, Florida, USA},
  pages     = {248--255},
  publisher = {{IEEE} Computer Society},
  year      = {2009},
  doi       = {10.1109/CVPR.2009.5206848},
  url       = {https://doi.org/10.1109/CVPR.2009.5206848}
}

@book{goodfellow2016deep,
  title     = {Deep Learning},
  author    = {Goodfellow, Ian and Bengio, Yoshua and Courville, Aaron},
  year      = {2016},
  publisher = {MIT Press}
}

@inproceedings{Krizhevsky2012AlexNet,
  author    = {Alex Krizhevsky and Ilya Sutskever and Geoffrey E. Hinton},
  title     = {ImageNet Classification with Deep Convolutional Neural Networks},
  booktitle = {Advances in Neural Information Processing Systems 25 (NeurIPS 2012)},
  editor    = {Peter L. Bartlett and Fernando C. N. Pereira and Christopher J. C. Burges and L{\'{e}}on Bottou and Kilian Q. Weinberger},
  pages     = {1106--1114},
  year      = {2012},
  url       = {https://proceedings.neurips.cc/paper/2012/hash/c399862d3b9d6b76c8436e924a68c45b-Abstract.html}
}

@article{Heinrich2012MIND,
  author  = {Heinrich, Mattias P. and Jenkinson, Mark and Bhushan, Manav and Matin, Tahreema and Gleeson, Fergus V. and Brady, Michael and Schnabel, Julia A.},
  title   = {MIND: Modality independent neighbourhood descriptor for multi-modal deformable registration},
  journal = {Medical Image Analysis},
  year    = {2012},
  month   = oct,
  volume  = {16},
  number  = {7},
  pages   = {1423--1435},
  doi     = {10.1016/j.media.2012.05.008},
  pmid    = {22722056},
  note    = {Epub 2012-05-31},
  url     = {https://pubmed.ncbi.nlm.nih.gov/22722056/}
}

@inproceedings{DeMan2007CatSim,
  author    = {De Man, Bruno and Basu, Samit and Chandra, Naveen and Dunham, Bruce and Edic, Pete and Iatrou, Maria and McOlash, Scott and Sainath, Paavana and Shaughnessy, Charlie and Tower, Brendon and Williams, Eugene},
  title     = {CatSim: a new computer assisted tomography simulation environment},
  booktitle = {Medical Imaging 2007: Physics of Medical Imaging},
  editor    = {Hsieh, Jiang and Flynn, Michael J.},
  series    = {Proceedings of SPIE},
  volume    = {6510},
  pages     = {65102G},
  year      = {2007},
  publisher = {SPIE},
  doi       = {10.1117/12.710713}
}
\bibliographystyle{spiejour}

\listoffigures
\listoftables

\end{spacing}
\end{document}